\documentclass[acmsmall,screen]{acmart}
\acmBooktitle{Code Review Automation Via Multi-task Federated LLM - An Empirical Study}
\AtBeginDocument{%
  \providecommand\BibTeX{{%
    \normalfont B\kern-0.5em{\scshape i\kern-0.25em b}\kern-0.8em\TeX}}}

\usepackage{float}
\usepackage{enumitem}
\usepackage{tcolorbox}
\usepackage{tabularx}
\usepackage{subcaption}
\usepackage{multirow}
\usepackage{colortbl}
\usepackage{array}
\usepackage[utf8]{inputenc}
\usepackage{booktabs} % For better looking tables
\usepackage{caption}
\usepackage{booktabs} % Required for nicer horizontal rules in tables
\usepackage{siunitx}
\usepackage{graphicx}
\usepackage{longtable}
\usepackage{amsmath}
\usepackage{ragged2e}

\begin{document}

%\title{Approaching Code Review Automation with Multi-task FedLLM training}

% \title{Code Review Automation, Multi-task and Federated LLM - An Empirical Study}

\title{Code Review Automation Via Multi-task Federated LLM - An Empirical Study}

% Simple approaches to Multi-task FedLLM training for Code Review Automation
% Split or No Split - Code Review Automation Multi-task FedLLM training
%\title{Code Review Automation with sequential Multi-task FedLLM training}
% Continual Multi-task FedLLM training for Code Review Automation
% Catastrophic forgetting in the continual Multi-task FedLLM training for Code Review Automation
% Empirical study towards privacy-aware Code Review Automation using Multi-task FedLLM
% Towards building Multi-task FedLLM: Demonstrated for Code Review Automation

\author{Jahnavi Kumar}
\affiliation{\textit{Research in Intelligent Software \& Human Analytics Lab}\\
Department of Computer Science and Engineering\\
Indian Institute of Technology Tirupati
\country{India}}
\email{cs22s503@iittp.ac.in}

\author{Sridhar Chimalakonda}
\affiliation{\textit{Research in Intelligent Software \& Human Analytics Lab}\\
Department of Computer Science and Engineering\\
Indian Institute of Technology Tirupati
\country{India}}
\email{ch@iittp.ac.in}
\renewcommand{\shortauthors}{Kumar and Chimalakonda}

\begin{abstract}
% Code Review is a crucial process before deploying code to production, as it validates the code, provides suggestions for improvements, and identifies corrections or errors such as missed edge cases. In projects with regular production releases, peer review efforts for code review remain high. Consequently, there have been significant efforts by software engineering (SE) researchers to automate the code review process. Previous works on code review automation approaches the task as three sequential sub-tasks: review necessity prediction, review comment generation, and code refinement. This study aims (i) to develop a multi-task model that addresses all the sub-tasks of Code review automation. (ii) Given the proprietary nature of code, we train the large language model (LLM) using federated learning (FL). 
Code review is a crucial process before deploying code to production, as it validates the code, provides suggestions for improvements, and identifies errors such as missed edge cases. In projects with regular production releases, the effort required for peer code-reviews remains high. Consequently, there has been significant interest from software engineering (SE) researchers in automating the code review process. Previous research on code review automation has typically approached the task as three independent sub-tasks: review necessity prediction, review comment generation, and code refinement. 

Our study attempts to (i) leverage the relationships between the sub-tasks of code review automation, by developing a multi-task model that addresses all tasks in an integrated manner, and (ii) increase model robustness on unseen data via collaborative large language model (LLM) modeling, while retaining the proprietary nature of code, by using federated learning (FL). The study explores five simple techniques for multi-task training, including two sequential methods, one parallel method, and two cumulative methods.

The results indicate that sequentially training a federated LLM (FedLLM) for our code review multi-task use case is less efficient in terms of time, computation, and performance metrics, compared to training separate models for each task. Because sequential training demonstrates catastrophic forgetting, alternatively cumulative fine-tuning for multi-task training performs better than training models for individual tasks. This study highlights the need for research focused on effective fine-tuning of multi-task FedLLMs for SE tasks.
% Code Review is a crucial process before deploying code to production, as it validates the code, provides suggestions for improvements, and identifies corrections or errors such as missed edge cases. In projects with regular production releases, peer review efforts for code review remain high. Consequently, there have been significant efforts by software engineering (SE) researchers to automate the code review process. In this paper, we build upon previous works on code review automation, approaching the task as three sequential sub-tasks: i) review necessity prediction, ii) review comment generation, and iii) code refinement. This study aims to develop a multi-task model that addresses all these sub-tasks. Given the proprietary nature of code, we train the large language model (LLM) using federated learning (FL). 
% The results indicate that sequential training a federated LLM (FedLLM) for our Code Review multi-task use case is time-consuming and less efficient, both in terms of time, computation, and performance metrics, compared to training separate models for each task. While results demonstrate catastrophic forgetting, this calls for research focused on the effective fine-tuning of multi-task FedLLMs for SE tasks. 

\end{abstract}



\maketitle

\section{Introduction}
\label{sec: Intro}
% Intro to Code Review
The code review task aids in identifying bugs early and improving code quality in terms of comprehension and performance. Developers conduct peer reviews of code changes before deploying them to the production environment to determine if improvements or corrections are necessary \cite{kononenko2016code}. If updates are required, the next steps involve identifying the specific updates and refining the code accordingly. However, code reviews demand significant effort from peer reviewers and are time-consuming. Most developers spend one to five hours per week on them \cite{eisty2022developers}, with Microsoft developers dedicating 15-25\% of their time, and open-source developers often spending even more \cite{morales2015code}. This has led researchers to explore automation solutions \cite{davila2021systematic, tufano2021towards, bacchelli2013expectations}.
% Models for Code review
Code analysis tools such as FindBugs were integrated to assist in code review process \cite{ayewah2007using, bacchelli2013expectations}. Later, researchers leveraged deep learning models, such as encoder-decoder transformers for code review comment generation, and refined-code review tasks where 16-31\% cases provided meaningful recommendations \cite{tufano2021towards}. With the growing modeling techniques, recent code review automation research \cite{lu2023llama, pornprasit2024fine} employ Large Language Models (LLMs). 

% Aim1: FedLLM
%ch 
However, the LLM based approaches for code review have multiple concerns: (i) proprietary code has to be shared with LLM, which may not be feasible in real-world scenarios, and (ii) most available LLMs are trained on open-source data for code review, limiting model's applicability to non-open source (private or closed) code review \cite{shanbhag2022exploring}. Consequently, companies and individuals are fine-tuning LLMs with their private data for specific tasks. 
\textit{This raises the question of whether they have \textbf{sufficient code data} to fine-tune LLMs individually}. In many companies, code is siloed. Hence, to build a robust fine-tuned LLM, the departments or companies could try collaborative modeling. Given the proprietary nature of code, where sharing code for collaborative learning is not feasible, Federated (fed) Learning (FL) \cite{bonawitz2017practical} could serve as a solution. FL is suitable as it allows sharing locally trained model weights (indirectly, the knowledge learned from private data) with participants/clients \cite{yang2024federated}.
There are limited works in software engineering (SE) that explore the application of FL to train LLMs for specific code tasks, such as code summarization \cite{kumar2024code}, code generation \cite{chen2024promise}; however, the applicability of FedLLM for code review automation tasks remains unexplored. \textit{The first aim of the study is to \textbf{explore FedLLM modeling} for code review automation}. 

% multiple related sub-tasks 
% Aim2: multi-task
Because of the complexity of the code review automation process, researchers have approached automating the code review process through a series of related tasks \cite{lu2023llama, li2022automating}. Li et al. \cite{li2022automating} approaches this automation with three main sub-tasks: (i) code review necessity prediction (RNP), which determines whether an input code patch (difference hunk) requires further inspection by outputting a Yes or No, (ii) code review comment generation (RCG), which provides feedback by producing code comments based on the input code patch, and (iii) code refinement (CR), which generates updated code based on the initial code and the generated review comments. The first two tasks focus on the reviewer’s perspective, assessing the need for a review and creating feedback, while the third task involves the developer's perspective, implementing the suggested changes.
The DISCOREV study \cite{ben2024improving} demonstrated that the three code review tasks are closely intertwined, and their interconnectedness can be leveraged for modeling code review, through cross-task knowledge distillation. This inspires us to build a multi-task model which implicitly takes advantage of the connectedness between these code review tasks. \textit{Our second aim is to \textbf{harness the inter-dependence} of code review tasks via multi-task training}.   

\begin{figure}[h]
    \centering
    \includegraphics[scale=0.65]{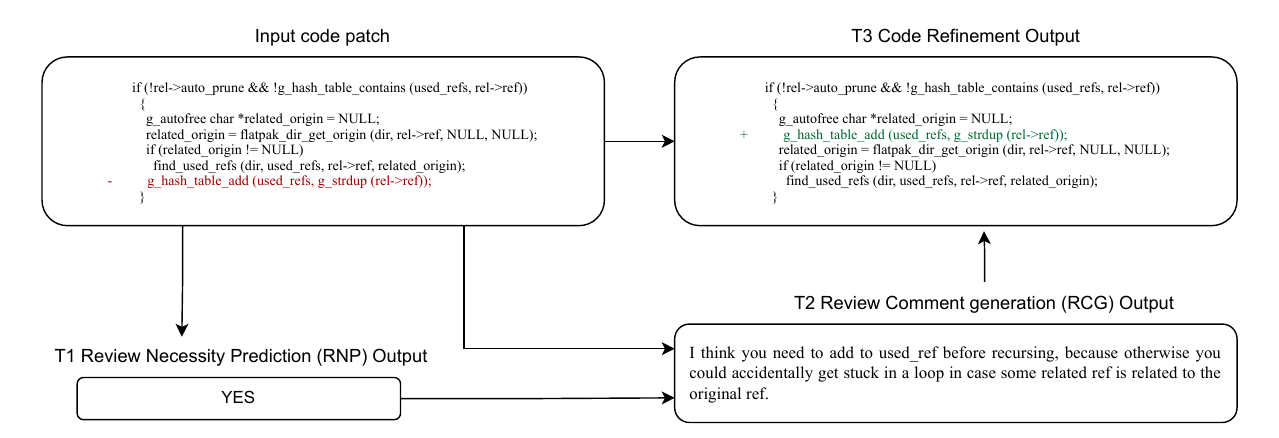}
    \caption{Example of Code Review with Task Workflow.}
    \label{fig:ProcessDiag}
\end{figure}

To summarize, the tasks involved in the code review automation process are interrelated, as illustrated by the example from the \texttt{CodeReviewer} dataset \cite{li2022automating} in Figure \ref{fig:ProcessDiag}. This study aims to implicitly utilize the relationships among these tasks to build a multi-task FedLLM that excels across all three introduced tasks. Inspired from the existing code review study \cite{lu2023llama} which uses LLaMA-family model, we chose to experiment with the recent version available at beginning of study, Meta's 8B LLaMA-3 model\footnote{LLaMA-3 release blog: \url{https://ai.meta.com/blog/meta-llama-3/}}. To maintain the privacy-aware aspect of this study, we avoided using any closed-source models \cite{sallou2024breaking}. Instead, we choose to experiment with an open source model for building task-wise and multi-task FedLLM for code review automation. 

We approach building multi-task FedLLM with five simple techniques: two sequential, one parallel and two cumulative. To validate that our multi-task FedLLM built for code review automation performs comparably or better than individual task models built with FedLLM, we first establish a baseline by creating individual task models in Research Question 1 (RQ1). We then proceed to evaluate the main objective of our study, which is the multi-task FedLLM, in Research Question 2 (RQ2).
The following research questions (RQs) will determine whether the multi-task FedLLM can achieve performance levels similar to those of task-specific individual FedLLMs:

\noindent\textbf{RQ1: Does federated fine-tuning with private code samples help build better models for each sub-task in Code Review Automation? (Homogeneous task)}
\begin{itemize}
    \item \noindent \textit{\textbf{Motivation:}} (i) The limited applicability of pre-trained LLMs built on open-source code to proprietary code review task. (ii) The lack of non-siloed data that can effectively facilitate the fine-tuning of LLMs for proprietary code review task.  

    \item \noindent \textit{\textbf{Experiment:}} We examine whether fine-tuning the LLM with our private data improves model inference on our specific dataset compared to the vanilla pre-trained LLM. In this context, we build separate models for three sub-tasks: review necessity prediction, review comment generation, and code refinement.

    \item \noindent \textit{\textbf{Key findings:}} Our results indicate that for review necessity prediction, and review comment generation, the FedLLM is particularly beneficial for low-data client. However, for code refinement, the FedLLM is found advantageous for both high-data and low-data clients.

    \item \noindent \textit{\textbf{Potential implication:}} Clients with limited code data could participate in privacy-aware collaborative training using FL to build a robust models capable of addressing code review tasks. Future research could investigate the integration of additional privacy-preserving techniques, such as homomorphic encryption, to enhance the security of collaborative FedLLM modeling. 
\end{itemize}
\noindent\textbf{RQ2: Can we build a single multi-task model that performs well across all three sub-tasks of Code Review Automation? (Heterogeneous tasks)}
\begin{itemize}
    \item \noindent \textit{\textbf{Motivation:}} The three code review tasks are closely intertwined, hence we want to harness their inter-dependencies implicitly via multi-task training. 
    % We believe the connectedness of sub-tasks could be advantageous in multi-task training, and private participants could try to build a single collaborative model together which can handle all the connected tasks robustly for all the clients. 

    \item \noindent \textit{\textbf{Experiment:}} We investigate whether the interconnectedness of these tasks can be leveraged to build a multi-task model that performs comparably or better across all code review sub-tasks when compared to the individual models developed in RQ1.

    \item \noindent \textit{\textbf{Key findings:}} Our findings indicate that sequential training of a FedLLM on the code review tasks results in catastrophic forgetting, making it an unstable approach. In contrast, the cumulative fine-tuning technique for the multi-task FedLLM demonstrates improved performance over individual-task models in the context of code review automation.

    \item \noindent \textit{\textbf{Potential implication:}} Training a multi-task FedLLM is advantageous for clients aiming to model related tasks such as tasks within the code review automation process. Future research could focus on exploring continual learning techniques to mitigate or reduce catastrophic forgetting during the training of multi-task FedLLMs.  
\end{itemize}

\textit{To the best of our knowledge, this is the first instance of approaching multi-task FedLLM in SE, specifically for code review automation. The study consumed 2600+ wall-clock hours on a server equipped with NVIDIA A100 40GB GPU.} 

\setlength{\fboxrule}{1.5pt}
\noindent\fcolorbox{yellow!50!black}{yellow!10!white}{%
    \parbox{\dimexpr\linewidth-2\fboxsep-2\fboxrule\relax}{%
        \strut Existing LLMs often struggle to effectively review unseen private code. This paper aims to:
        \begin{enumerate}[label=\roman*)]
        \item Experiment privacy-aware collaborative fine-tuning of LLM using FL for code review automation tasks, and 
        \item Leverage the inter-dependencies among code review automation tasks (review necessity prediction, review comment generation, and code refinement) by building a multi-task FedLLM. 
        \end{enumerate}
Our findings demonstrate that this multi-task FedLLM performs comparably to or even outperforms individual-task FedLLMs across all the three chosen code review tasks. 
        \strut
    }%
}

% <TBD contributions TBD>

% The paper is structured as follows: Background about training the model in Section \ref{sec: bkgrnd}, followed by the experimental design \ref{sec: exp}, methodology in Section \ref{sec: method}, and evaluation in Section \ref{sec:eval}. The discussion, threats to validation, related work are presented in Sections \ref{sec: Discuss}, \ref{sec: Threats}, \ref{sec: Related}, respectively. The paper concludes with future work in section \ref{sec: ConclFuture}. 

\section{Background}
\label{sec: bkgrnd}
% \noindent\textbf{Code Review: }
% \noindent 
% Developers conduct peer reviews of code changes before deploying them to the production environment to determine if improvements or corrections are necessary \cite{kononenko2016code}. If updates are required, the next steps involve identifying the specific updates and refining the code accordingly. The automated approach to the code review process \cite{tufano2021towards, bacchelli2013expectations}, inspired by Li et al. \cite{li2022automating}, encompasses multiple related tasks: (i) code review necessity prediction, with the code patch (difference hunk) as input and a Yes/No output, (ii) code review comment generation, with the code patch as input and code comments as output, and (iii) code refinement, where the updated code is generated based on the inputted source code and review comments. The first two tasks are from the reviewer’s perspective, while the third is from the developer’s perspective.

% <LLMs + PEFT>
\noindent\textbf{Fine-tuning LLM: }
\noindent 
Large Language Models (LLMs), typically having billions of parameters, are trained on vast amounts of data, enabling them to handle various language tasks across different domains such as translation, summarization, and question answering \cite{ozkaya2023application}. To excel in a specific task or domain, LLMs can be further trained (fine-tuned) with particular datasets \cite{ozkaya2023application, pornprasit2024fine}. Fine-tuning LLMs is computationally demanding and time-consuming, leading researchers to explore Parameter Efficient Fine Tuning (PEFT), which fine-tunes only a select number of model weights, and performs as comparable as fine-tuning all parameters \cite{h2024technical}. Among PEFT methods, Low Rank Adaptation (LoRA) \cite{hu2021lora} is popular \cite{fakih2024llm4plc, lu2023llama}, focusing on training representative-weights of the model. The change in the weight matrix \( \Delta W \)  matrix is decomposed into two lower-rank matrices \( A \) and \( B \), which are trained during the fine-tuning process. Their matrix multiplication \( A \cdot B \) is then added to the original weight matrix \( W \) to obtain the new model weights \( W_{1} \), as described in Equation \ref{eqn:LoRA}. 
\begin{equation}
\label{eqn:LoRA}
W_{1} = W + \Delta W = W + A \cdot B
\end{equation}
where ${W_{1}, W} \in \mathbb{R}^{d \times k}$, $A \in \mathbb{R}^{d \times r}$, $B \in \mathbb{R}^{r \times k}$, and  the rank $r \ll \min(d, k)$.
\\

\noindent\textbf{Federated Learning:}
\noindent
Federated Learning (FL) is a collaborative mechanism for training model where clients share their locally trained model weights with a central server (or other clients) \cite{nguyen2021federated}. The server aggregates these weights and sends the updated model back to the clients. This process is repeated for multiple fed rounds, with each new round using the updated model as the base. FL is a privacy-aware method because clients do not share their data; instead, they share the knowledge learned from their data through model weights \cite{bonawitz2017practical}.
\begin{figure}[h]
    \centering
    \includegraphics[scale=0.58]{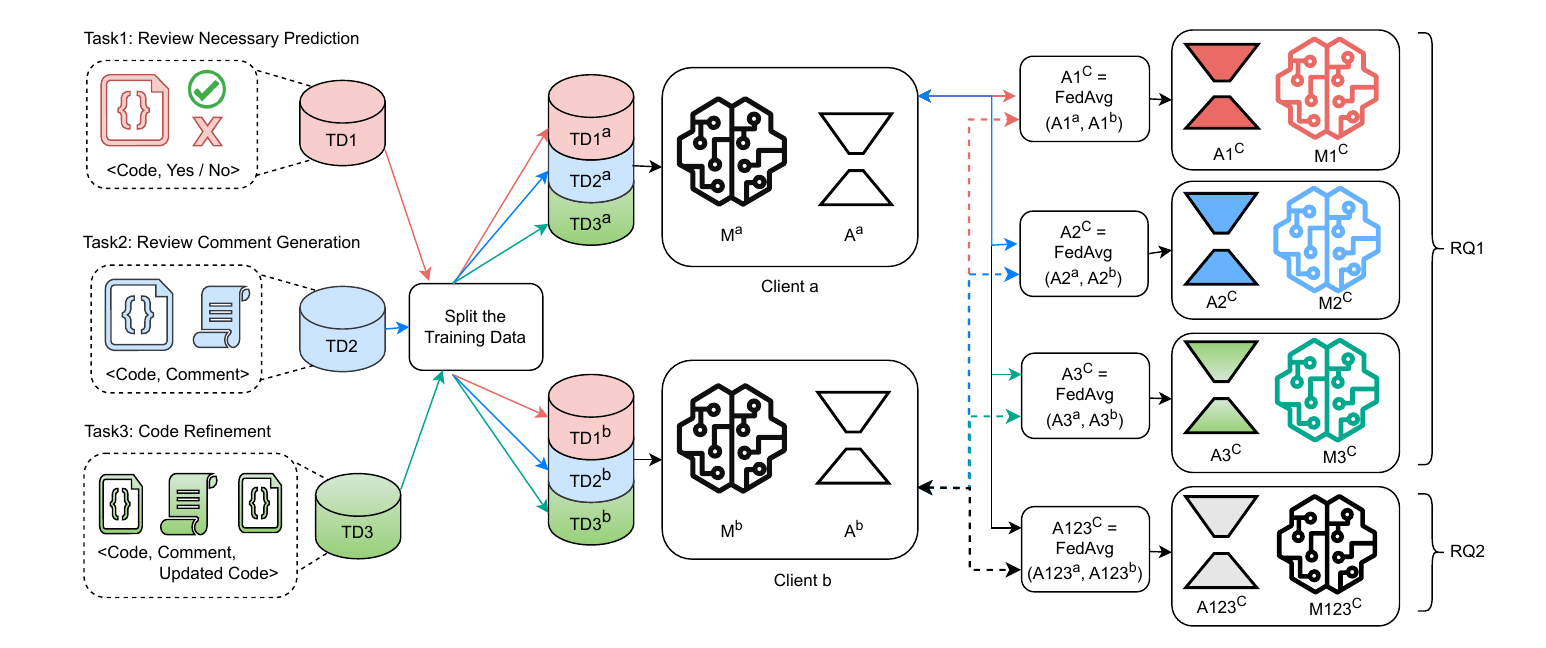}
    \caption{Overall architecture diagram, illustrating Tasks 1, 2, and 3 and Clients a and b.}
    \label{fig:MethodDiag}
\end{figure}
 
This process adapted from existing work \cite{kumar2024code} can be observed in Figure \ref{fig:MethodDiag}, where Client \(a\) fine-tunes model \(M\) on its training data \(TD^a\) using LoRA to obtain the adapter \(A^a\). Similarly, Client \(b\) fine-tunes model \(M\) on \(TD^b\) to obtain adapter \(A^b\). These LoRA adapters \(A^a\) and \(A^b\) are then shared with the central server, which aggregates the adapter weights and returns the updated adapter \(A^C\). The clients merge this updated adapter with the model to get the updated model \(M^t\) for round \(t\), which is evaluated using test data metrics before initiating the next round \(t+1\). This process continues for a predetermined number of fed rounds \(T\) or until the clients achieve optimal results, indicated by minimal improvements or performance above a set threshold.

\section{Experimental Design}
\label{sec: exp}
The proposed approach is explained in this section. 
% <Proposed Approach>
\subsection{Fine-tuning LLaMA-3 for Code Review}
Inspired by the LLaMAReviewer paper \cite{lu2023llama}, this study employs LoRA to fine-tune the LLaMA-3 8B model, an open-source LLM, for code review tasks. Each client performs supervised fine-tuning on its own data, after which the server aggregates and shares the adapter for the next round. We utilize the multi-language dataset from the CodeReviewer (CRer) paper \cite{li2022automating}, which includes data from nine popular programming languages on GitHub: C, C++, C\#, Go, Java, JavaScript, PHP, Python, and Ruby. The pre-processed CRer dataset is divided into three subsets, corresponding to each task in the code review automation process.
\\

\noindent\textit{\textbf{Pre-processing of dataset:}} 
For the federated learning simulation, we ensure a heterogeneous data distribution, meaning no projects are shared between clients. The pre-processing guarantees that the \texttt{client-0 train}, \texttt{client-1 train}, and \texttt{test} datasets are heterogeneous, as the study aims to assess the robustness of trained models on unseen project data.

a) The \texttt{train} dataset for T1 had project information missing in several rows, therefore we designated the \texttt{train} dataset as client-0's dataset and the validation dataset for as client-1's dataset. This approach was similarly applied for T2 and T3.

b) While the training dataset is distinct from the validation and test datasets, there were shared projects in the validation and test datasets. To address this, we merged the \texttt{validation} and \texttt{test} datasets, selecting the top 50\% of projects with the highest data points as the \texttt{new-valid} dataset (for client-1) and designating the remaining projects as the \texttt{new-test} dataset.

In summary, the CodeReviewer dataset comprises three subsets for three different tasks. We utilize all datasets, assigning the \texttt{train}, \texttt{new-valid}, and \texttt{new-test} subsets as \texttt{client-0 train}, \texttt{client-1 train}, and \texttt{test} datasets, respectively, for all three tasks. The final project selections for each client are documented in the replication package\footnote{Final Dataset project split used in study: \url{https://osf.io/ztnhk/files/osfstorage/67442c97742ad8612f7a4b28}}. This version maintains the original meaning while improving readability and reducing redundancy. 
\\

\noindent\textit{\textbf{Preparation of dataset for FL:}}
We selected two clients with a training dataset ratio of 3:1, specifically 19,500 to 6,500 samples, following the methodology in \cite{shanbhag2022exploring}, which employs a fixed number of clients for simulation. This ratio allows us to simulate a scenario where clients have unequal data amounts while still enabling knowledge sharing. We believe collaboration is more appealing when clients have a reasonable amount of data relative to each other; extreme ratios such as 100:1, may reduce interest in collaboration. Since real-world scenarios rarely achieve even a 1:1 ratio, we chose a 3:1 ratio to enhance replicability and practicality.

\begin{figure}[h]
    \centering
    \begin{subfigure}[b]{0.49\linewidth} 
        \centering
        \includegraphics[width=\linewidth]{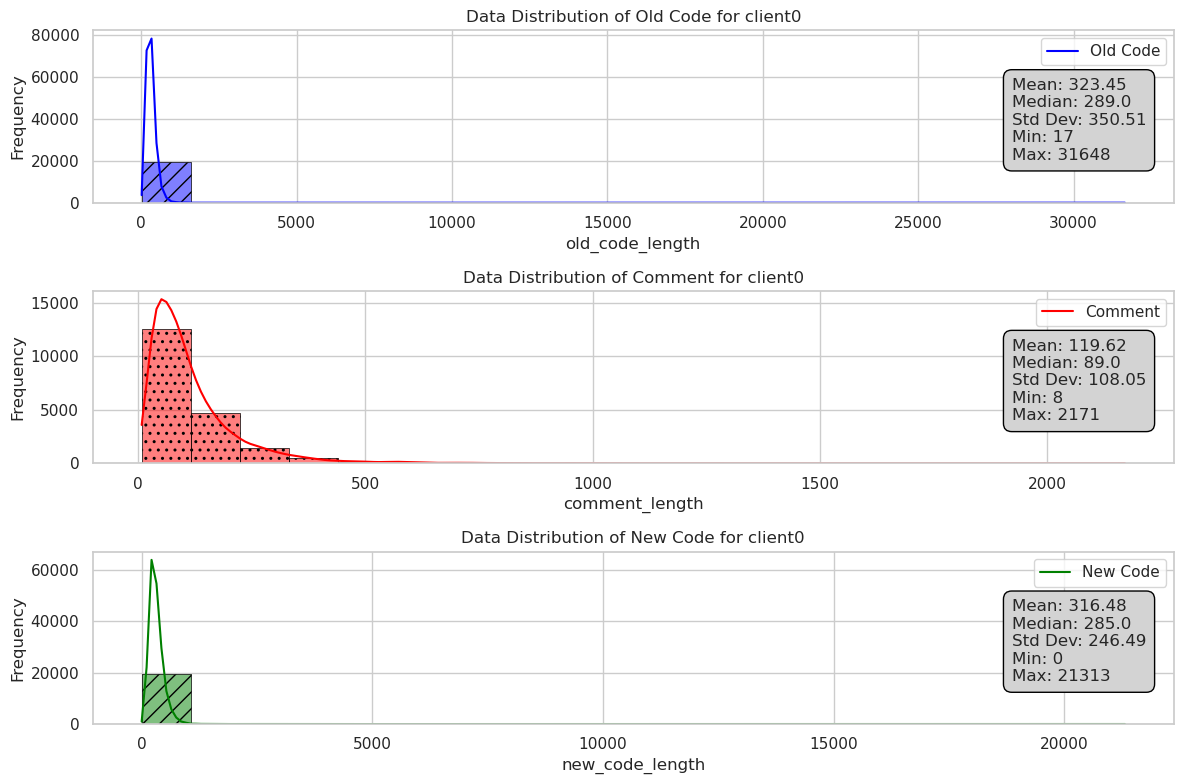} 
        \caption{Data Distribution of client a}
        \label{fig:client0_DSDistri}
    \end{subfigure}
    \hfill
    \begin{subfigure}[b]{0.49\linewidth} 
        \centering
        \includegraphics[width=\linewidth]{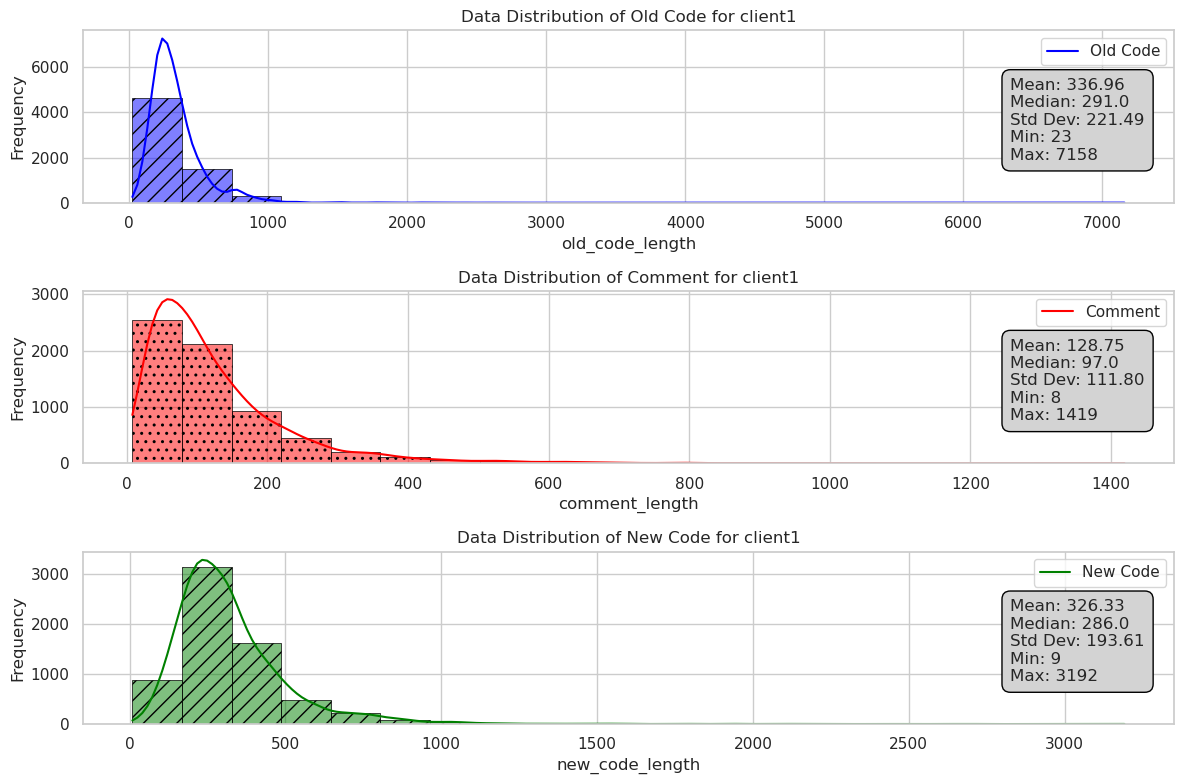} 
        \caption{Data Distribution of client b}
        \label{fig:client1_DSDistri}
    \end{subfigure}
    \caption{Data Distribution of clients a and b}
    \label{fig:clients_DSDistri}
\end{figure}

The available dataset is categorized into ten groups based on patch length. From each category, we select one data point per iteration until the required number is reached. For Task 1, we ensure an equal distribution of yes and no labels, totaling the required amount. We selected 26,000 training data points—consistent with the optimal dataset size for SE code task identified by \cite{kumar2024code} and used by Snowflake \footnote{Snowflake fine-tuning LLM on roughly 26k SQL data-points: \url{https://www.snowflake.com/en/blog/meta-code-llama-testing/}}—and 2,000 test data points (keeping the patch length below 5,000 to avoid quantization and out-of-memory errors, where length refers to the number of characters in the code/comment) for each task, as detailed in Table \ref{tab:chosen_dataset}, where the `\texttt{Total Train}' is the sum of the data from clients a and b. For Task 1, we selected 13,000 data points that require no review and 13,000 that do require review. Figure \ref{fig:clients_DSDistri} illustrates the heterogeneity among client datasets. 

\begin{table}[hbt]
\captionsetup{skip=0pt}
\centering
\caption{Chosen dataset. T: Total-available, C: Chosen\vspace{0.5em}}
\label{tab:chosen_dataset}
\captionsetup[subtable]{justification=centering,singlelinecheck=false}

\begin{subtable}[t]{\linewidth}
\centering
\caption{Review Necessity Prediction (RNP) (T1)}
\label{tab:subtask1}
\small
\begin{tabular}{|l|c|c|c|c|c|c|c|c|}
\hline
\rowcolor{gray!30}
\textbf{} & \multicolumn{2}{c|}{\textbf{Client a}} & \multicolumn{2}{c|}{\textbf{Client b}} & \multicolumn{2}{c|}{\textbf{Total Train}} & \multicolumn{2}{c|}{\textbf{Test}} \\ \hline
\rowcolor{gray!20}
 & \textbf{YES} & \textbf{NO} & \textbf{YES} & \textbf{NO} & \textbf{YES} & \textbf{NO} & \textbf{YES} & \textbf{NO} \\ \hline
T & 132918 & 132918 & 25609 & 26906 & 158527 & 159824 & 5643 & 4346 \\ \hline
C & 9750 & 9750 & 3250 & 3250 & 13000 & 13000 & 1000 & 1000 \\ \hline
\end{tabular}
\end{subtable}

\vspace{1em}

\begin{minipage}{0.49\linewidth}
\centering
\begin{subtable}[t]{\linewidth}
\centering
\caption{Review Comment Generation (RCG) (T2)}
\label{tab:subtask2}
\small
\begin{tabular}{|l|l|l|l|l|}
\hline
\rowcolor{gray!30}
\textbf{} & \textbf{Client a} & \textbf{Client b} & \textbf{Total Train} & \textbf{Test} \\ \hline
T & 117739 & 9971 & 127710 & 3817 \\ \hline
C & 19500 & 6500 & 26000 & 2000 \\ \hline
\end{tabular}
\end{subtable}
\end{minipage}
\hfill
\begin{minipage}{0.49\linewidth}
\centering
\begin{subtable}[t]{\linewidth}
\centering
\caption{Code Refinement (CR) (T3)}
\label{tab:subtask3}
\small
\begin{tabular}{|l|l|l|l|l|}
\hline
\rowcolor{gray!30}
\textbf{} & \textbf{Client a} & \textbf{Client b} & \textbf{Total Train} & \textbf{Test} \\ \hline
T & 150406 & 22398 & 172804 & 3809 \\ \hline
C & 19500 & 6500 & 26000 & 2000 \\ \hline
\end{tabular}
\end{subtable}
\end{minipage}

\end{table}

\noindent\textit{\textbf{Prompting and evaluation details:}}
We utilize the training and inference prompts from \cite{lu2023llama} for supervised fine-tuning, omitting language names due to their minimal impact on performance, resulting in more generic prompts. For all tasks, the input includes the difference hunk, while Task 3 also incorporates the review as input. The outputs are as follows: Task 1 produces a yes/no response, Task 2 generates a natural language (NL) comment, and Task 3 outputs updated programming language (PL) code. 

The performance of the classification task (T1), Review Necessity Prediction, is evaluated using precision, recall, and F1 score \cite{lu2023llama}. For regression tasks T2 and T3, the base paper LLaMAReviewer \cite{lu2023llama} employs BLEU-4 with macro-averaging. However, since BLEU primarily measures n-gram lexical similarity between generated and referenced output at the corpus level, it is less effective for individual sentences. Therefore, we adopt Corpus-BLEU (C-BLEU) based on the micro-averaging approach proposed by Papineni et al. \cite{papineni2002bleu}. In addition, in line with existing research \cite{niu2022spt, kumar2024code}, we utilize METEOR (Metric for Evaluation of Translation with Explicit Ordering), which incorporates synonyms before checking n-gram similarity, and ROUGE-L (Recall-Oriented Understudy for Gisting Evaluation - Longest Common Subsequence), which measures the longest common subsequence between generated and reference outputs.
% METEOR and ROUGE-L are particularly suitable metrics for evaluating the NL generation task (T2). 

\subsection{Non-federated approach baseline}
We train the selected LLM on the complete code review dataset \(TD\) \cite{lu2023llama}, which is the merged data from all clients (\(TD^a\) and \(TD^b\)). This scenario represents the best-case scenario where all clients' private data is accessible for creating a combined dataset to train a central model. However, this approach is neither ideal nor practical for situations involving private or siloed code. 

% \begin{figure}[h]
%     \centering
%     \includegraphics[scale=0.255]{CR_Figures/test_DSDistri.png}
%     \caption{Data Distribution of test set}
%     \label{fig:test_DSDistri}
% \end{figure}

% <<Prompt, Instr> for Federated approach>
% \subsection{Prompt for Supervised fine-tuning}
\subsection{Hyper-parameter configuration}
Experiments for hyper-parameter search of were conducted similarly to methodology outlined in \cite{kumar2024code} utilizing a random selection of 10,000 balanced data points from the training dataset. These experiments were performed on an NVIDIA A100 40GB GPU. For the fine-tuning process, we set lora\_alpha to 16 and applied a dropout probability of 0.1 to the LoRA layers. Training was performed over one epoch, utilizing half-precision (fp16) training to optimize computational efficiency. The per-device batch size for training was set to 2, with gradient accumulation steps at 1, and the maximum gradient norm clipped at 0.3. The initial learning rate was 0.0003, optimized using the AdamW optimizer, with a weight decay of 0.001. The learning rate followed a cosine scheduler, and the warm-up ratio was set to 0.03. These settings closely follow the configurations in the base paper \cite{lu2023llama}, with modifications to the learning rate scheduler and warm-up ratio inspired by \cite{kumar2024code}.

% <RQs>
\subsection{RQ implementation design}
% <Evaluation>
% <Implementation details, configurations>
\noindent\textbf{RQ1: Does federated fine-tuning with private code samples help build better models for each sub-task in Code Review Automation? (Task homogeneity)}

\noindent We construct models \( M^1 \), \( M^2 \), and \( M^3 \) for tasks \( T_1 \), \( T_2 \), and \( T_3 \) respectively. In the initial round (round zero), adapters are initialized on top of vanilla LLaMA 3 following the LoRA methodology \cite{hu2021lora}. In round \( t \) for task \( T_1 \), client \( a \) trains \( M^1_C(t-1) \) on \( TD^1_a \) and shares the trained adapter \( A^a(t) \) with the central server. Concurrently, client \( b \) trains \( M^1_C(t-1) \) on \( TD^1_b \) and shares the trained adapter \( A^b(t) \) with the central server. The central server employs the popular federated aggregation method, FedAvg \cite{shanbhag2022exploring, li2019convergence}, on the LoRA adapters \( A^a(t) \) and \( A^b(t) \), and distributes the resultant \( A^C(t) \) to the clients. Clients then merge the received \( A^C(t) \) with their model and consider it as the base model for the subsequent round. This process repeats for \( T \) rounds for task \( T_1 \). The same procedure is applied for \( T \) rounds for tasks \( T_2 \) and \( T_3 \) as well.

In this study, we run for 1 to \( T \) rounds and select the best model among all \( T \) models. However, in a real-world scenario, clients can terminate the process when they believe they have reached the expected performance or when further training yields negligible improvements. The resultant models \( M1^C \), \( M2^C \), and \( M3^C \) represented as the top three models in Figure \ref{fig:MethodDiag} are used by all clients for inference on tasks \( T1 \), \( T2 \), and \( T3 \) respectively.
\\

\noindent\textbf{RQ2: Can we build a single multi-task model that performs well for all three sub-tasks of Code Review Automation? (Task heterogeneity)}

\noindent The base paper \cite{lu2023llama} trains an adapter for each task, which can be merged with a shared LLM. This approach leverages the LoRA adapters to store a single model for all three tasks. Inspired by this, we aim to train a single model using FL for all three tasks. 
% While FL facilitates privacy-preserving collaboration, it requires multiple rounds and may not achieve the same performance as a centrally trained model with access to all data.
However in FL, as the number of rounds increases, the base models for each task tend to diverge. This divergence makes it challenging to build a single shared model with different task adapters. Consequently, we train the single model on all the tasks either sequentially, parallelly or cumulatively to make it multi-task model. With this aim, we propose five approaches which includes two sequential, one parallel and two cumulative training approaches. 
% Given the differing training methods required for classification and regression tasks, we avoid simultaneous training on \(T_1\), \(T_2\), and \(T_3\).

\subsubsection{\textbf{Sequential approaches for multi-task training}}
% Instead, we adopt a sequential training approach. 
We first train a model \(M\) on \(T_1\) to obtain \(M_1\). Next, we train \(M_1\) on \(T_2\) to get \(M_2\). Finally, we train \(M_2\) on \(T_3\) to obtain \(M_3\). This sequential training process aligns with the actual code review automation sequence, with the expectation that the knowledge gained from \(T_1\) will benefit \(T_2\), and the combined knowledge from \(T_1\) and \(T_2\) will benefit \(T_3\).
In this aspect, we try two approaches: 
% \begin{itemize}

\noindent\textit{\textbf{Federated Task-Of-Clients (FedTOC)}}: 
We perform FL involving the clients and server for \(T_1\) first to build \(M_1\). Here, all the clients finalize that the \(T_1\) training is completed. Next, the clients and server together consider \(M_1\) as the base and initiate FL for \(T_2\) to build \(M_{12}\). The clients declare that their \(T_2\) training is completed and start training on \(M_{12}\) for \(T_3\). The resultant \(M_{123}\) is trained on all the tasks using the knowledge from all the clients, making it the final model that all clients rely on for all tasks. The approach is named so because FL between clients happens first, then is repeated for all tasks  as shown in Equation \ref{eqn:FedTaskOfClients}. 
\begin{equation}
\label{eqn:FedTaskOfClients}
% M_{123}^C = \text{Fed}(M_{12}^a, M_{12}^b), \quad \text{where} \quad M_{12}^C = \text{Fed}(M_{1}^a, M_{1}^b), \quad M_{1}^C = \text{Fed}(M_{1}^a, M_{1}^b)
M_{(123)}^C = \text{Fed}(M_{(12)3}^a, M_{(12)3}^b), \quad \text{where} \quad M_{(12)}^C = \text{Fed}(M_{(1)2}^a, M_{(1)2}^b), \quad M_{(1)}^C = \text{Fed}(M_{1}^a, M_{1}^b)
\end{equation} 
    
\noindent\textit{\textbf{Federated Client-Of-Tasks (FedCOT)}}:
Every client trains the model \(M\) on \(T_1\) to get \(M_1\), then trains \(M_1\) on \(T_2\) to get \(M_{12}\), and finally trains \(M_{12}\) on \(T_3\) to get \(M_{123}\). The clients then share the trained adapter corresponding to \(M_{123}\) with the server. The server aggregates and shares the aggregated \(M_{123}\). This process is repeated for \(T\) rounds or until the clients declare that they have reached the required performance or that further training does not yield significant improvements. The final \(M_{123}\) serves as the final model all clients rely on for all tasks. This approach is named so because training is continued by clients individually for all tasks first, then FL is applied between clients as shown in Equation \ref{eqn:FedClientOfTasks}. 
\begin{equation}
\label{eqn:FedClientOfTasks}
M_{123}^C = \text{Fed}(M_{123}^a, M_{123}^b) 
% (\Sigma_{\text{T1, T2, T3}}) 
\end{equation}

\subsubsection{\textbf{Parallel approach for multi-task training}}
We train a model \(M\) on \(T_1\), \(T_2\) and \(T_3\) in parallel to obtain \(M_{123}\). In this aspect, we try below approach:

\noindent\textit{\textbf{Federated Clients-And-Tasks (FedCAT)}}:
In this approach, we perform FL where each client trains separately on all three tasks \(T_1\), \(T_2\), and \(T_3\). Each client first completes training on the tasks independently, producing models \(M_{1}^a, M_{1}^b, M_{2}^a, M_{2}^b, M_{3}^a,\) and \(M_{3}^b\)
% for tasks \(T_1, T_2,\) and \(T_3\), respectively
. Once the individual models are trained, the FL process involves combining all these models across tasks and clients to generate a unified model, \(M_{\text{final}}\) as shown in Equation \ref{eqn:FedClientsAndTasks}. This final model integrates the knowledge from each client and each task, resulting in a shared model suitable for all tasks and clients. 
% The approach is named so because the FL processes both cross-client and cross-task training simultaneously, as depicted in Equation \ref{eqn:FedClientsAndTasks}.
% We perform FL involving the clients and server for \(T_1\) first to build \(M_1\). Here, all the clients finalize that the \(T_1\) training is completed. Next, the clients and server together consider \(M_1\) as the base and initiate FL for \(T_2\) to build \(M_{12}\). The clients declare that their \(T_2\) training is completed and start training on \(M_{12}\) for \(T_3\). The resultant \(M_{123}\) is trained on all the tasks using the knowledge from all the clients, making it the final model that all clients rely on for all tasks. The approach is named so because FL between clients happens first, then is repeated for all tasks  as shown in Equation \ref{eqn:FedClientsAndTasks}. 
\begin{equation}
\label{eqn:FedClientsAndTasks}
M_{\text{final}} = \text{Fed}(M_{T}^X \mid T \in \{1, 2, 3\}, X \in \{a, b\})
% M_{\text{final}} = \text{Fed}(M_{1}^a, M_{1}^b, M_{2}^a, M_{2}^b, M_{3}^a, M_{3}^b)
% M_{\text{final}} = \frac{1}{3} \sum_{\text{T} \in \{1, 2, 3\}} \text{Fed}(M_{T}^a, M_{T}^b)
% \Sigma_{\text{Y belongto T1, T2, T3}}^{\text{X belongsto a, b}} (\text{Fed}(M_{Y}^X))
% \Sigma_{\text{T belongsto 1, 2, 3}} (\text{Fed}(M_{T}^a, M_{T}^b))
\end{equation} 

\subsubsection{\textbf{Cumulative approaches for multi-task training}}
We cumulatively train a model \(M\) using a combined dataset of \(T_1 + T_2 + T_3\). Within this framework, we explore two specific approaches:
% We cumulatively train a model M on T1 + T2 + T3 dataset. In this aspect, we try below two approaches:
\noindent\textit{\textbf{Federated Cumulative Fine-Tuning (FedCFT)}}:
We propose applying the Cumulative Fine-Tuning (CFT) technique for training, where each client first trains a model on its own training data \(TD\), a random mix of data points from \(T_1, T_2,\) and \(T_3\). After the individual training, FL is performed between the client models. This one-shot (cumulative) approach trains on all tasks simultaneously, and hence the name FedCFT.
% We propose using the well-known Cumulative Fine-Tuning technique to train each client then FL between the client models. While using CFT, each client trains on it's own training data TD which is random mixture of T1 + T2 + T3 datapoints. This technique uses one-shot (cumulative) approach on training for all tasks and hence the name FedCFT. 

\noindent\textit{\textbf{Federated Cumulative Fine-Tuning - separate for Regression (FedCFT-reg)}}:
This technique is an enhanced version of the FedCFT approach. Considering the different training methods required for classification and regression tasks, FedCFT\_reg involves training two separate models using the FedCFT approach: one for the classification task \(T_1\) and another for the regression tasks \(T_2\) and \(T_3\). Because we train the regression tasks together, separately from the classification task, the technique is named FedCFT-reg.
% This technique is an improvised version of FedCFT technique. Given the differing training methods required for classification and regression tasks, this approach trains two separate models using FedCFT approach: one for classification task \(T_1\) and other for regression tasks \(T_2\), and \(T_3\). Because the regression tasks are trained together separately from the classification tasks, the technique is named FedCFT\_reg.  

\section{Methodology}
\label{sec: method}
In this section, we detail the methodology used to address the research questions. The initial step involves a hyper-parameter search before implementing the main methodology.

\subsection{RQ0: Hyper-parameter Search}
When we refer to sharing adapters, we mean sharing the low-rank matrices of LoRA. For each of the 32 layers of chosen LLM, we target selected linear matrices such as query (q), key (k), value (v), and output (o). The total number of sub-components in the state dictionary (representative of the adapters) that clients share is calculated as \(32 \times \text{(\#target-modules)} \times (2\) low-rank matrices), as only these updated LoRA weights are shared while the rest of the weights remain frozen. We denote the weight matrices of q, k, v, and o as \(W_q\), \(W_k\), \(W_v\), and \(W_o\) respectively.

This study performs a hyper-parameter search for target modules (matrices) and the LoRA rank \(r\) as described in \cite{kumar2024code}. In the target module search, we aim for trainable parameters of LLaMA-3 to be less than approximately 8M. For the rank search, we perform an unrestricted search until an out-of-memory error occurs. 

\noindent\textit{\textbf{Target module \(W\) search}} is performed on multiple dimensions (dim) of the target modules \cite{kumar2024code}:
\begin{itemize}
    \item \textbf{4 Dim}: (\(W_q, W_v, W_k, W_o\)) with rank 8
    \item \textbf{2 Dim}: (\(W_q, W_o\)), (\(W_k, W_o\)), (\(W_v, W_o\)), (\(W_q, W_k\)), (\(W_q, W_v\)), (\(W_k, W_v\)) with rank 16
    \item \textbf{1 Dim}: (\(W_q\)), (\(W_v\)), (\(W_k\)), (\(W_o\)) with rank 32
\end{itemize}
\noindent We select the two sets of target modules that perform the best for each task. For \(T_1\), we select (\(W_q\)) and (\(W_k, W_v\)). For \(T_2\), we select (\(W_q, W_o\)) and (\(W_v\)). For \(T_3\), we select (\(W_o, W_q\)) and (\(W_o\)). Additionally, as a third pick, we also choose the 4-dimension matrix (\(W_q, W_v, W_k, W_o\)) for all tasks.
% and second best

\textit{\textbf{LoRA rank \(r\) search}} is performed for for powers of 2 between 1 and 128, specifically 1, 2, 4, 8, 16, 32, 64, and 128, on the chosen target modules \cite{kumar2024code}. The hyper-parameters for \(T_1\) are finalized based on the F1 score, while those for \(T_2\) and \(T_3\) are chosen based on the best performance in at least two metrics among C-BLEU, METEOR, and ROUGE-L.
For \(T_1\), (\(W_k, W_v\)) performs best with ranks \(r=64\) and \(r=8\), with \(r=8\) being optimal. For \(T_2\), (\(W_v\)) performs best with \(r=8\). For \(T_3\), (\(W_q, W_v, W_k, W_o\)) and (\(W_q, W_o\)) perform best with \(r=16\), with (\(W_q, W_o\)) at \(r=16\) being optimal.

The finalized hyper-parameters are shown in Table \ref{tab:task_target_rank}. 
\begin{table}[ht]
\centering
\sloppy
\caption{Target modules and rank for different tasks.}
\label{tab:task_target_rank}
\begin{tabular}{|l|l|l|}
\hline
\rowcolor{gray!30}
\textbf{Task \(T\)} & \textbf{Target modules \(W\)} & \textbf{Rank \(r\)} \\ \hline
T1 & \(W_k\), \(W_v\) & 8 \\ \hline
T2 & \(W_v\) & 8 \\ \hline
T3 & \(W_q\), \(W_o\) & 16 \\ \hline
\end{tabular}
\end{table}
\subsection{RQ1: Individual Models for Each Task}
For each task, we evaluate the chosen LLM, LLaMA-3, on the respective task dataset. This evaluation is conducted using the vanilla model, or the 0th Federated round evaluation. Subsequently, clients train the model on their respective portions of the dataset and share the adapters with the server. The server aggregates these adapters and shares the aggregated adapter back with the clients. This aggregated adapter is merged with the 0th round base model for the 1st round evaluation. The newly merged model becomes the base for the next round, and this process continues for \( T \) rounds. In our experiments, we set \( T \) to 20, as suggested by \cite{kumar2024code}.
 
For reference, we also train a model using a non-federated (central) approach. It is important to note that the federated approach requires multiple rounds to achieve results comparable to the central model. Although we do not use the central model in our research question evaluations, we include its metrics to illustrate the trade-off in performance when opting for a privacy-aware federated approach over a central model. Our experiment calculated all of the metrics in the 0-1 range, however they are presented in percentages in the paper (tables).
 
\textbf{\textit{For \( T_1 \), Review Necessity Prediction}}, a classification task, we consider requiring a review as class 1 and not requiring a review as class 0. Table \ref{tab:combined_round_metrics} presents the precision, recall, and F1 scores for \( T_1 \), comparing the central (non-federated) model, the vanilla model (Fed 0), and federated models from rounds 1 to 20. The chosen projection matrices (\( W_k, W_v \)) with rank 8 result in the model being 0.0326\% trainable with 2.6M trainable parameters. 
% Although our experiment calculated all metrics in the 0-1 range, they are represented as percentages in the table.

\begin{table}[ht]
\centering
\sloppy
\caption{Round-wise Metrics for T1, T2, and T3 in percentages (\%)}
\label{tab:combined_round_metrics}
\begin{tabular}{|l|S[table-format=2.2]|S[table-format=2.2]|S[table-format=2.2]|S[table-format=2.2]|S[table-format=2.2]|S[table-format=2.2]|S[table-format=2.2]|S[table-format=2.2]|S[table-format=2.2]|}
\hline
\rowcolor{gray!30}
\textbf{Round} & \multicolumn{3}{c|}{\textbf{\footnotesize T1 (RNP)}} & \multicolumn{3}{c|}{\textbf{\footnotesize T2 (RCG)}} & \multicolumn{3}{c|}{\textbf{\footnotesize T3 (CR)}} \\ \cline{2-10}
\rowcolor{gray!30}
& \scriptsize \textbf{PRECISION} & \scriptsize \textbf{RECALL} & \scriptsize \textbf{F1} & \scriptsize \textbf{C-BLEU} & \scriptsize \textbf{METEOR} & \scriptsize \textbf{ROUGE-L} & \scriptsize \textbf{C-BLEU} & \scriptsize \textbf{METEOR} & \scriptsize \textbf{ROUGE-L} \\ \hline
\rowcolor{gray!15}
Central & 48.851 & 63.800 & 55.334 & 0.568 & 9.624 & 11.389 & 74.749 & 84.336 & 88.394 \\ \hline
0 & 53.989 & 29.100 & 37.817 & 0.110 & 8.840 & 7.706 & 12.607 & 44.878 & 34.451 \\ \hline
\multicolumn{1}{|l|}{\cellcolor{gray!20}\bfseries 1} & 
\multicolumn{1}{S[table-format=2.2]|}{\cellcolor{gray!20}\bfseries 49.141} & 
\multicolumn{1}{S[table-format=2.2]|}{\cellcolor{gray!20}\bfseries 62.900} & 
\multicolumn{1}{S[table-format=2.2]|}{\cellcolor{gray!20}\bfseries 55.175} & 
\multicolumn{1}{S[table-format=2.2]|}{\cellcolor{gray!20}\bfseries 0.612} & 
\multicolumn{1}{S[table-format=2.2]|}{\cellcolor{gray!20}\bfseries 9.245} & 
\multicolumn{1}{S[table-format=2.2]|}{\cellcolor{gray!20}\bfseries 10.850} & 72.627 & 84.005 & 86.277 \\ \hline
2 & 54.121 & 39.400 & 45.602 & 0.463 & 8.582 & 10.529 & 73.492 & 84.402 & 86.964 \\ \hline
3 & 64.829 & 24.700 & 35.771 & 0.440 & 8.462 & 10.482 & 74.825 & 84.623 & 87.282 \\ \hline
4 & 66.301 & 24.200 & 35.458 & 0.317 & 8.312 & 10.426 & 74.878 & 84.575 & 87.367 \\ \hline
5 & 68.120 & 25.000 & 36.576 & 0.356 & 8.295 & 10.365 & 74.684 & 84.533 & 87.391 \\ \hline
6 & 66.418 & 26.700 & 38.088 & 0.350 & 8.286 & 10.272 & 74.499 & 84.417 & 87.324 \\ \hline
7 & 66.933 & 25.100 & 36.509 & 0.413 & 8.389 & 10.298 & 74.574 & 84.499 & 87.395 \\ \hline
\cellcolor{gray!20}\textbf{8} & 70.317 & 24.400 & 36.229 & 0.463 & 8.525 & 10.387 & \multicolumn{1}{S[table-format=2.2]|}{\cellcolor{gray!20}\bfseries 74.583} & \multicolumn{1}{S[table-format=2.2]|}{\cellcolor{gray!20}\bfseries 84.542} & \multicolumn{1}{S[table-format=2.2]|}{\cellcolor{gray!20}\bfseries 87.472} \\ \hline
9 & 70.588 & 25.200 & 37.141 & 0.461 & 8.571 & 10.469 & 74.479 & 84.494 & 87.393 \\ \hline
10 & 72.372 & 24.100 & 36.159 & 0.509 & 8.472 & 10.516 & 74.448 & 84.513 & 87.409 \\ \hline
11 & 73.457 & 23.800 & 35.952 & 0.505 & 8.541 & 10.514 & 74.432 & 84.435 & 87.252 \\ \hline
12 & 73.846 & 24.000 & 36.226 & 0.475 & 8.419 & 10.535 & 74.298 & 84.323 & 87.257 \\ \hline
13 & 76.282 & 23.800 & 36.280 & 0.474 & 8.430 & 10.605 & 74.290 & 84.344 & 87.151 \\ \hline
14 & 74.823 & 21.100 & 32.917 & 0.575 & 8.497 & 10.588 & 74.336 & 84.291 & 87.134 \\ \hline
15 & 75.932 & 22.400 & 34.595 & 0.543 & 8.440 & 10.537 & 74.198 & 84.205 & 87.034 \\ \hline
16 & 75.987 & 23.100 & 35.429 & 0.515 & 8.364 & 10.433 & 74.312 & 84.300 & 87.137 \\ \hline
17 & 76.667 & 23.000 & 35.385 & 0.451 & 8.422 & 10.455 & 74.308 & 84.281 & 87.089 \\ \hline
18 & 75.399 & 23.600 & 35.948 & 0.454 & 8.415 & 10.460 & 74.316 & 84.291 & 87.153 \\ \hline
19 & 75.229 & 24.600 & 37.076 & 0.451 & 8.390 & 10.430 & 74.284 & 84.252 & 87.092 \\ \hline
20 & 74.426 & 22.700 & 34.789 & 0.525 & 8.509 & 10.516 & 74.250 & 84.241 & 87.052 \\ \hline
\end{tabular}
\end{table}

\textbf{\textit{For \( T_2 \), Review Comment Generation}}, a regression task, we present the C-BLEU, METEOR, and ROUGE-L metrics for the central model, the vanilla model (Fed 0), and the federated models evaluated after each round (Round 1 to  20) in Table \ref{tab:combined_round_metrics}. The chosen projection matrix (\( W_v \)) with rank 8 result in the model being 0.0163\% trainable with 1.3M trainable parameters. 
% Although 

\textbf{\textit{For \( T_3 \), Code Refinement}}, we present the evaluated performance of the central model, the vanilla model (Fed 0), and the federated models in Table \ref{tab:combined_round_metrics}. We use the C-BLEU, METEOR, and ROUGE-L metrics for evaluation, as it is a regression task. The chosen projection matrices (\( W_q, W_o \)) with rank 16 result in the model being 0.104\% trainable with 8.4M trainable parameters. 
% Although our experiment calculated all metrics in the 0-1 range, they are represented as percentages in the table.

\noindent The evaluation and the analysis of RQ1 results is in the next section, Section \ref{sec:eval}.
%<Lets give the tables for all the 3 tasks> <Lets also say the evaluation and the analysis is in next section>

\subsection{RQ2: Multi-task Model for All Tasks}
In code review automation, tasks are typically performed sequentially, leading us to initially mimic this approach by training models sequentially. However, we also explored parallel training, where the model is trained on all tasks simultaneously. To address the limitations of both techniques, we adopted cumulative fine-tuning technique. While we propose these techniques, future research could explore other continuous fine-tuning techniques to further enhance model performance.

\textbf{Fed Task-Of-Clients}: This method involves the sequential training of clients for each task. Initially, clients collaboratively perform federated training for \( T_1 \), followed by federated training for \( T_2 \), and finally for \( T_3 \). It is important to note that each training phase builds sequentially upon the previous one. 
Each client initially trains the chosen LLM \( M \) on their data for \( T_1 \) using LoRA with target modules (\( W_k, W_o \)) and a rank of 8. The clients select the federated model at round 1 as the final \( M_1 \), as it demonstrates the best performance, as shown in Table \ref{tab:combined_round_metrics}. Subsequently, clients train \( M_1 \) for the \( T_2 \) task using LoRA with target module (\( W_v \)) and a rank of 8. They determine that the optimal performance is achieved at federated round 8, which they declare as \( M_{12} \). Finally, the clients train \( M_{12} \) for the \( T_3 \) task using LoRA with target modules (\( W_q, W_o \)) and a rank of 16. The optimal performance is identified at federated round 2, resulting in the final model \( M_{123} \).

\textbf{Fed Client-Of-Tasks}: This method involves sequential training of individual clients for the \( T_1 \), \( T_2 \), and \( T_3 \) tasks, followed by model exchange for FL. It is important to note that within each client, training is performed sequentially based on the previous task's model. In this continuous training method, to avoid contradictory training progress from previous tasks, we maintain consistency in the target modules by using all four matrices (\( W_q, W_v, W_k, W_o \)) with a rank of 8 for all tasks.
Each client initially trains the chosen LLM \( M \) on their data for \( T_1 \) to obtain their own \( M_1 \). Note that the \( M_1 \) models of different clients are distinct. Then, each client trains their \( M_1 \) for \( T_2 \) to obtain their own \( M_{12} \). Finally, each client trains their \( M_{12} \) for \( T_3 \) to obtain their own \( M_{123} \). The clients then share their respective \( M_{123} \) models with the server for FL and receive back the aggregated final \( M_{123} \), which all clients will use as the base for the next round. This process is repeated for \( T \) rounds, and in our experiments, we have set \( T \) to 20. 

\textbf{Fed Clients-And-Tasks}: This method involves parallel training of individual clients on tasks \(T_1\), \(T_2\), and \(T_3\), followed by model exchange for FL. Within each client, training for each task is conducted simultaneously, ensuring consistency across target modules by using all four matrices (\(W_q, W_v, W_k, W_o\)) with a rank of 8. Initially, each client trains the chosen LLM \(M\) on their local data to obtain models \(M_1\), \(M_2\), and \(M_3\) for the respective tasks. These models are distinct across clients. Afterward, clients share their models with the server, which aggregates them to produce a final model \(M_{123}\). This aggregated model is then used as the base for subsequent rounds of training. In this approach, for each client, all tasks are weighted equally during aggregation. This process is repeated for \( T \) rounds, and in our experiments, we have set \( T \) to 20.
% This method involves parallel training of individual clients for the \( T_1 \), \( T_2 \), and \( T_3 \) tasks separately, followed by model exchange for FL. It is important to note that within each client, training for each task is performed in parallel. In this parallel training method, we maintain consistency in the target modules by using all four matrices (\( W_q, W_v, W_k, W_o \)) with a rank of 8 for all tasks.
% Each client initially trains the chosen LLM \( M \) on their data for \( T_1 \), \( T_2 \), and \( T_3 \) in parallel to obtain their own \( M_1 \), \( M_2 \), \( M_3 \). Note that the \( M_1 \) models of different clients are distinct. Then, the clients share their respective \( M_1 \), \( M_2 \), \( M_3 \) models with the server for FL and receive back the aggregated final \( M_{123} \), which all clients will use as the base for the next round. In this paper, all the tasks are given equal weightage when aggregating. This process is repeated for \( T \) rounds, and in our experiments, we have set \( T \) to 20.

\textbf{Fed Cumulative Fine-Tuning}: This method involves training individual clients on a cumulative dataset \(TD_1 + TD_2 + TD_3\), followed by model exchange for FL to obtain \(M_{123}\). Within each client, the randomly-mixed cumulative dataset is trained using LoRA with a rank of 8, targeting the modules \(W_q, W_v, W_k,\) and \(W_o\). The aggregated model from each round is then used as the base for subsequent rounds. This process is repeated for \(T = 20\) rounds.
% This method involves training of individual clients on cumulative dataset \(TD_1\) + \(TD_2\) + \(TD_3\), followed by model exchange for FL to obtain \(M_{123}\). Within each client, the randomly-mixed cumulative dataset is  trained with LoRA of rank 8 targeting the modules \(W_q, W_v, W_k, W_o\). This aggregated model at each round is used as the base for subsequent rounds. This process is repeated for \( T \) = 20 rounds.

\textbf{Fed Cumulative Fine-Tuning - separate for Regression}: 
The method involves training two separate models using the FedCFT methodology: one for the classification task and another for the regression tasks. For the regression model, individual clients are trained on the cumulative dataset \(TD_2 + TD_3\), which includes the regression tasks \(T_2\) and \(T_3\), followed by model exchange for FL to obtain \(M_{23}\). The classification task RNP \(T_1\), we use the individual-task model trained for \(T_1\) in RQ1, instead of re-training the same for evaluation.
% The method involves training two models, one for classification task and other for regression tasks. For the regression model, we train individual clients on cumulative dataset \(TD_2\) + \(TD_3\) of regression tasks \( T_2 \) and \( T_3 \), followed by model exchange for FL to obtain \(M_{23}\). We then follow the FedCFT methodology for training both the classification and regression model. The classification task RNP \( T_1 \) is the individual-task model trained for T1 in RQ1, we do not retrain and use the same for evaluation. 

\noindent The evaluation and analysis of RQ2 results are presented in the next section, Section \ref{sec:eval}.

\section{Evaluation}
\label{sec:eval}
This section evaluates and compares the task-wise individual models with the multi-task model.
\subsection{RQ1: Individual Models for Each Task}
We build individual models for each task. From Table \ref{tab:combined_round_metrics}, we can see that the central model performs better because it can access and train on the entire dataset at once, rather than through multiple federated rounds. However, we choose FL because it offers privacy. 

\textbf{\textit{For \( T_1 \), Review Necessity Prediction}}, we observe from Table \ref{tab:combined_round_metrics} that the federated model at round 1 performs the best, and any additional rounds of training degrade the model's performance. Hence, we select FedBEST round as 1, with precision, recall, and F1 scores of 49\%, 62\%, and 55\% respectively. This indicates that federated fine-tuning improves the precision, recall, and F1 of the vanilla model by 4\%, 33\%, and 18\% respectively. It also demonstrates that additional federated rounds do not necessarily enhance the model.
\textbf{\textit{For \( T_2 \), Review Comment Generation}}, we observe from Table \ref{tab:combined_round_metrics} that the federated model at round 1 performs the best, and any additional rounds of training degrade the model's performance. Hence, we select FedBEST round as 1, with C-BLEU, METEOR, and ROUGE-L metrics of 0.6\%, 9.2\%, and 10.9\% respectively. This shows that federated fine-tuning improves the BLEU, METEOR, and ROUGE-L metrics of the vanilla model by 0.5\%, 0.4\%, and 3.2\% respectively. It also shows that additional federated rounds do not necessarily mean a better model, as the scores decline with additional rounds after round 1.
\textbf{\textit{For \( T_3 \), Code Refinement}}, we observe from Table \ref{tab:combined_round_metrics} that the federated model at round 8 performs the best, and any additional rounds of training degrade the model's performance. Hence, we select FedBEST round as 8, with C-BLEU, METEOR, and ROUGE-L metrics of 74.58\%, 84.54\%, and 87.47\% respectively. This shows that fed fine-tuning improves the BLEU, METEOR, and ROUGE-L metrics of the vanilla model by 62\%, 40\%, and 53\% respectively. It also shows that additional federated rounds do not necessarily mean a better model, as the scores decline with additional rounds after round 8.

Compared to all the individual models, FL has the most significant impact on \( T_3 \), Code Refinement, as we observe a substantial increase in the model's performance compared to the vanilla non-fine-tuned model. The FL training has also improved the performance of \( T_1 \), Review Necessity Prediction, to a lesser extent. However, FL did not significantly enhance the performance of \( T_2 \), Review Comment Generation. It is noteworthy that the chosen raw LLM exhibited weak performance with low metric values on the selected dataset for \( T_2 \) specifically, which explains the marginal increase in the metric values. It is important to note that these results are specific to this dataset and model. Using a different model for the \( T_2 \) task might yield better results.

\begin{table}[ht]
\centering
\small
\sloppy
\caption{Evaluation Metrics of Individual Models. Metrics are in percentages (\%).}
\label{tab:RQ1evaluation_metrics}
\begin{tabular}{|l|S[table-format=2.2]|S[table-format=2.2]|S[table-format=2.2]|S[table-format=2.2]|S[table-format=2.2]|S[table-format=2.2]|S[table-format=2.2]|S[table-format=2.2]|S[table-format=2.2]|}
\hline
\rowcolor{gray!30}
\textbf{Model} & \multicolumn{9}{c|}{\textbf{Task Metrics}} \\ \hline
& \multicolumn{3}{c|}{\footnotesize \textbf{T1 (RNP)}} & \multicolumn{3}{c|}{\footnotesize \textbf{T2 (RCG)}} & \multicolumn{3}{c|}{\footnotesize \textbf{T3 (CR)}} \\ \hline
\footnotesize \textbf{BEST} & \multicolumn{3}{c|}{1} & \multicolumn{3}{c|}{1} & \multicolumn{3}{c|}{8} \\ \hline
\rowcolor{gray!30}
& \scriptsize \textbf{PRECISION} & \scriptsize \textbf{RECALL} & \scriptsize \textbf{F1} & \scriptsize \textbf{C-BLEU} & \scriptsize \textbf{METEOR} & \scriptsize \textbf{ROUGE-L} & \scriptsize \textbf{C-BLEU} & \scriptsize \textbf{METEOR} & \scriptsize \textbf{ROUGE-L} \\ \hline
\footnotesize \textbf{Vanilla} & 53.989 & 29.100 & 37.817 & 0.110 & 8.840 & 7.706 & 12.607 & 44.878 & 34.451 \\ \hline
\footnotesize \textbf{Central} & 48.851 & \cellcolor{gray!20}63.800 & \cellcolor{gray!20}55.334 & 0.568 & \cellcolor{gray!20}9.624 & \cellcolor{gray!20}11.389 & \cellcolor{gray!20}74.749 & \cellcolor{gray!20}84.336 & \cellcolor{gray!20}88.394 \\ \hline
\footnotesize \textbf{Client a@1} & \cellcolor{gray!20}48.884 & \cellcolor{gray!20}63.500 & \cellcolor{gray!20}55.241 & 0.527 & \cellcolor{gray!20}9.268 & \cellcolor{gray!20}10.975 & 69.488 & 83.656 & 86.144 \\ \hline
\footnotesize \textbf{Client b@1} & \cellcolor{gray!20}64.567 & 8.200 & 14.552 & 0.464 & 7.460 & 9.826 & 64.612 & 81.581 & 83.856 \\ \hline
\footnotesize \textbf{Fed@1} & 49.141 & 62.900 & 55.175 & 0.612 & 9.245 & 10.850 & 72.627 & 84.005 & 86.277 \\ \hline
\footnotesize \textbf{Fed@BEST} & 49.141 & 62.900 & 55.175 & \cellcolor{gray!20}0.612 & 9.245 & 10.850 & \cellcolor{gray!20}74.583 & \cellcolor{gray!20}84.542 & \cellcolor{gray!20}87.472 \\ \hline
\end{tabular}
\end{table}

For all tasks \( T_1 \), \( T_2 \), and \( T_3 \), Table \ref{tab:RQ1evaluation_metrics} presents a comparison of the vanilla, central (non-federated), clients' models after fine-tuning once on their own data, the federated model after the first round, and the federated model at the BEST round, with BEST being 1, 1, and 8 for \( T_1 \), \( T_2 \), and \( T_3 \), respectively.
To illustrate this with examples, we display the input and outputs of the vanilla model, federated model, and central model in tables. Table \ref{tab:T2_outputs} presents anecdotal examples for \( T_2 \), where both the FedBEST and central models generate comments similar to the ground truth, which the pre-trained model fails to do. Table \ref{tab:T3_outputs} presents anecdotal examples for \( T_3 \), where the central model and FedBEST generates code identical to the ground truth, whereas the pre-trained model fails. Since \( T_1 \) is a classification task, we do not provide examples. 
% in this paper. 

\textit{For \( T_1 \) and \( T_2 \)}, the performance order is \textit{Central > Client a@1 > Fed1 = FedBEST > Client b@1 > Pre-trained}. Client b@1 performs poorly, indicating that federated learning is particularly beneficial for the low-resource Client b. This could be due to the test data being more closely related to Client a's data and the fact that Client a has significantly more data.
\textit{For \( T_3 \)}, the performance order is \textit{Central > FedBEST > Fed1 > Client a@1 > Client b@1 > Pre-trained}. This  demonstrates that federated learning is beneficial for both high-data and low-data clients.

\vspace{5pt}
\noindent\colorbox{yellow!50!black}{\parbox{\dimexpr\linewidth-2\fboxsep}{\color{yellow!10!white}\textbf{RQ1 Summary}}}
\noindent\fcolorbox{yellow!50!black}{yellow!10!white}{%
    \parbox{\dimexpr\linewidth-2\fboxsep-2\fboxrule\relax}{%
        \strut This study simulates a non-open source scenario with multiple clients and federated fine-tunes the 8B LLaMA-3 model using A100 GPU with 40GB RAM for code review automation tasks. For Review Necessity Prediction (T1), a classification task, and Review Comment Generation (T2), a natural language generation task, we found that the FedLLM is particularly beneficial for low-data client. However, for Code Refinement (T3), a programming language generation task, FedLLM proves advantageous for both high-data and low-data clients.
    }%
}
\vspace{5pt}
% <Explain the RQ1 table>
% Because T1 is classification task, we do not show any examples in this paper. Table \ref{tab:T2_outputs}, and Table shows anecdotal example for T2 and T3. We show the input, and outputs of vanilla model, Federated model, central model.   

\begin{table*}[ht]
\centering
\sloppy
\caption{Input and Model Outputs for T2 Review Comment Generation}
\label{tab:T2_outputs}
\begin{tabular}{|>{\raggedright\arraybackslash}m{2cm}|>{\noindent\justifying\arraybackslash}p{11cm}|}
\hline
\textbf{Input} &\footnotesize \begin{verbatim}
@@ -116,10 +116,16 @@ namespace MvvmCross.Binding.Bindings.Target.Construction
{    return factory;}
+ var implementedInterfaces = type.GetTypeInfo().ImplementedInterfaces;
+ foreach (var implementedInterface in implementedInterfaces)
+ {
+    factory = this.FindSpecificFactory(implementedInterface, name);
+    if (factory != null) return factory;
+ }
var baseType = type.GetTypeInfo().BaseType;
if (baseType != null)
    return this.FindSpecificFactory(baseType, name);
return null;
\end{verbatim} 
\\ \hline
\rowcolor{gray!30}
\textbf{Model} & \textbf{Output} \\ \hline
\footnotesize \textbf{Ground Truth} & \footnotesize I'm not too familiar with this code but shouldn't this be done after the base type? My gut says that we should check the base class before any interfaces that may be implemented. \\ \hline
\footnotesize \textbf{Pre-trained} & \footnotesize The code provided is a method for finding a specific factory based on a type and a name. The method first checks if the type has implemented interfaces, and if so, it iterates through the implemented interfaces to find a specific factory.
The code is well-structured and easy to follow. However, there are a few areas that could be improved:
% \begin{itemize}
%     \item The method is quite long and could be broken down into smaller, more manageable methods.
%     \item The method uses a lot of nested if statements, which can make the code harder to read and maintain. Consider using a more structured approach, such as using a switch statement or creating a separate method for each possible outcome.
%     \item The method does not handle errors well. If an error occurs, the method will simply return null, which may not be the desired behavior. Consider adding error handling to the method, such as using try-catch blocks or returning a specific error code.
% \end{itemize}
* The method is quite long and could be broken down into smaller, more manageable methods.
* The method uses a lot of nested if statements, which can make the code harder to read and maintain. Consider using a more structured approach, such as using a switch statement or creating a separate method for each possible outcome.
* The method does not handle errors well. If an error occurs, the method will simply return null, which may not be the desired behavior. Consider adding error handling to the method, such as using try-catch blocks or returning a specific error code.
Overall, the code is well-structured and easy to follow. However, there are a few areas that could be improved to make the code more robust and maintainable. \\ \hline
\footnotesize \textbf{Central} & \footnotesize I think this is a good idea, but I think we should also check for interfaces that are not implemented by the type itself, but by its base types. \\ \hline
\footnotesize \textbf{FedBEST} & \footnotesize I think this is a good idea, but I think it's a bit too much. We should probably only do this for interfaces that are explicitly implemented by the type. \\ \hline
\footnotesize \textbf{FedCFT} & \footnotesize think this is a good idea, but I think we should also check for `IMvxValueConverter` and `IMvxValueConverter<T>` interfaces. \\ \hline
\footnotesize \textbf{FedCFT-reg} & \footnotesize I think this is a good idea, but I think we should also add a check for `type.IsInterface` to avoid the case where we have a class that implements an interface. \\ \hline
\end{tabular}
\end{table*}

\begin{table}[ht]
\centering
\sloppy
\caption{Input and Model Outputs for T3 Code Refinement}
\label{tab:T3_outputs}
\begin{tabular}{|>{\raggedright\arraybackslash}m{2.2cm}|>{\raggedright\arraybackslash}p{10.8cm}|}
\hline
\textbf{Input} & \footnotesize\begin{verbatim}Comment:'Should not mention schnorr yet on BTC'
 Code: '  * @param {number} signature.sigtype
  * @param {PublicKey} signature.publicKey
  * @param {Signature} signature.signature
- * @param {String} signingMethod - 'ecdsa' or 'schnorr' to sign transaction
  * @return {Transaction} this, for chaining
  */Transaction.prototype.applySignature = function(signature, signingMethod) {
\end{verbatim}
\\ \hline
\rowcolor{gray!30}
\textbf{Model} & \textbf{Output} \\ \hline
\footnotesize \textbf{Ground Truth, Central, FedBEST, FedCFT, FedCFT-reg} & \footnotesize\begin{verbatim}  * @param {number} signature.sigtype
  * @param {PublicKey} signature.publicKey
  * @param {Signature} signature.signature
+ * @param {String} signingMethod - 'ecdsa' to sign transaction
  * @return {Transaction} this, for chaining
  */Transaction.prototype.applySignature = function(signature, signingMethod) {
\end{verbatim}
\\ \hline
\footnotesize \textbf{Pre-trained} & \footnotesize\begin{verbatim}e is the refined code based on the provided code review comment:
```Transaction.prototype.applySignature = function(signature, signingMethod) {
  if (signingMethod ==='schnorr') {
    // Add Schnorr signature logic here
  }
  // Rest of the method implementation
};```
\end{verbatim}
% \\ \hline
% \textbf{Central} & \begin{verbatim}
%      }
%      },
%      friendlyId: {
% +      presence: true
%      },
%      firstName: {
%        presence: true
% \end{verbatim}
\\ \hline
\end{tabular}
\end{table}
\subsection{RQ2: Multi-task Model for All Tasks}
In this section, we evaluate the five methods implemented in this study as explained in Section \ref{sec: method}: 

\noindent\textit{\textbf{FedTOC}}: The technique trains the model \( M_{123} \), where clients collaboratively train on tasks \( T_1 \), \( T_2 \), and \( T_3 \) sequentially. Table \ref{tab:FedTaskOfClients_metrics} shows the comparative performance of the intermediate models \( M_{1} \) and \( M_{12} \) with the final model \( M_{123} \).
Pre-training the model \( M_{123} \) on tasks \( T_1 \) and \( T_2 \) helps improve performance in terms of C-BLEU and METEOR metrics when fine-tuned for \( T_3 \). Specifically, pre-training increases C-BLEU and METEOR by 0.1\% and 0.5\% respectively, while ROUGE-L declines by 0.9\%. However, pre-training on \( T_1 \) does not improve performance on \( T_2 \).

The final \( M_{123} \) model suffers from catastrophic forgetting \cite{zhai2024investigating}, that is the model losses the ability to infer for previously trained tasks when trained on new tasks. This is evidenced by model's poor performance on \( T_1 \) and \( T_2 \) when compared to individual models trained in RQ1. This could be because of the diverse nature of the tasks in Code Review Automation, including classification (\( T_1 \)), natural language generation (\( T_2 \)) and programming language generation (\( T_3 \)). The \( T_1 \) classification task shows 0 values for all metrics. For \( T_2 \), a review (NL) generation task, the model \( M_{123} \) frequently generates code (PL) instead of natural language, including for the input example shown in Table \ref{tab:T2_outputs}. For \( T_3 \) input shown in Table \ref{tab:T3_outputs}, output is same as input code, which shows model performs poorly. 

\begin{table*}[ht]
\centering
\sloppy
\caption{Performance Metrics for FedTaskOfClients FedLLM (in \%)}
\label{tab:FedTaskOfClients_metrics}
\begin{tabular}{|l|>{\raggedleft\arraybackslash}p{0.5cm}|S[table-format=2.2]|S[table-format=2.2]|S[table-format=2.2]|S[table-format=2.2]|>{\raggedleft\arraybackslash}p{0.9cm}|S[table-format=2.2]|S[table-format=2.2]|>{\raggedleft\arraybackslash}p{0.9cm}|S[table-format=2.2]|}
\hline
\rowcolor{gray!30}
\scriptsize\textbf{Model} & \multicolumn{10}{c|}{\footnotesize\textbf{Task Metrics}} \\ \hline
\rowcolor{gray!30}
\footnotesize\textbf{} & & \multicolumn{3}{c|}{\footnotesize \textbf{T1 (RNP)}} & \multicolumn{3}{c|}{\footnotesize \textbf{T2 (RCG)}} & \multicolumn{3}{c|}{\footnotesize \textbf{T3 (CR)}} \\ \hline
\rowcolor{gray!30}
\footnotesize \textbf{}  & \scriptsize \textbf{BEST} & \scriptsize \textbf{PRECISION} & \scriptsize \textbf{RECALL} & \scriptsize \textbf{F1} & \scriptsize \textbf{C-BLEU} & \scriptsize \textbf{METEOR} & \scriptsize \textbf{ROUGE-L} & \scriptsize \textbf{C-BLEU} & \scriptsize \textbf{METEOR} & \scriptsize \textbf{ROUGE-L} \\ \hline
\textbf{\( M_{1} \)} & 1 & 49.141 & 62.900 & 55.175 & 0.000 & 0.000 & 0.000 & 0.000 & 0.000 & 0.000 \\ \hline
\textbf{\( M_{12} \)} & 8 & 0.000 & 0.000 & 0.000 & 0.655 & 8.218 & 10.324 & 0.141 & 1.239 & 2.564 \\ \hline
\rowcolor{gray!13}
\textbf{\( M_{123} \)} & 2 & 0.000 & 0.000 & 0.000 & 0.315 & 2.928 & 3.179 & 74.883 & 84.840 & 87.427 \\ \hline
\end{tabular}
\end{table*}

\noindent\textit{\textbf{FedCOT}}: Trains the model \( M_{123} \), where clients train sequentially on tasks \( T_1 \), \( T_2 \), and \( T_3 \), followed by FL. Table \ref{tab:FedClientOfTasks_metrics} shows the performance of the final model \( M_{123} \) for tasks \( T_1 \), \( T_2 \), and \( T_3 \). However, it is evident that Fed performs optimally at different rounds for each task: round 18 for \( T_1 \), round 2 for \( T_2 \), and round 1 for \( T_3 \). This indicates that additional Fed rounds degrade performance for \( T_2 \) and \( T_3 \).  
The final \( M_{123} \) model of this method performs better than the individual-task model for \( T_3 \), however performs poorly for \( T_1 \) and \( T_2 \) compared to the individual models trained in RQ1. It shows that this technique also suffers from catastrophic forgetting \cite{zhai2024investigating}. For this method, a single BEST round could not be determined for all tasks, highlighting the instability of this multi-task training approach for FedLLM, hence we haven't shown anecdotal examples.
% <method 2 we couldn't agree on the FedBEST round>
\begin{table}[ht]
\centering
\sloppy
\caption{Performance Metrics for FedClientOfTasks FedLLM (in \%)}
\label{tab:FedClientOfTasks_metrics}
\begin{tabular}{|l|S[table-format=2.2]|S[table-format=2.2]|S[table-format=2.2]|S[table-format=2.2]|S[table-format=2.2]|S[table-format=2.2]|S[table-format=2.2]|S[table-format=2.2]|S[table-format=2.2]|}
\hline
\rowcolor{gray!30}
\textbf{Model} & \multicolumn{9}{c|}{\textbf{Task Metrics}} \\ \hline
\rowcolor{gray!30}
\textbf{} & \multicolumn{3}{c|}{\footnotesize \textbf{T1 (RNP)}} & \multicolumn{3}{c|}{\footnotesize \textbf{T2 (RCG)}} & \multicolumn{3}{c|}{\footnotesize \textbf{T3 (CR)}} \\ \hline
\scriptsize \textbf{BEST} & \multicolumn{3}{c|}{18} & \multicolumn{3}{c|}{2} & \multicolumn{3}{c|}{1} \\ \hline
\rowcolor{gray!30}
\textbf{}  & \scriptsize \textbf{PRECISION} & \scriptsize \textbf{RECALL} & \scriptsize \textbf{F1} & \scriptsize \textbf{C-BLEU} & \scriptsize \textbf{METEOR} & \scriptsize \textbf{ROUGE-L} & \scriptsize \textbf{C-BLEU} & \scriptsize \textbf{METEOR} & \scriptsize \textbf{ROUGE-L} \\ \hline
\scriptsize \textbf{FedBEST} & 39.450 & 4.300 & 7.755 & 0.216 & 3.424 & 3.325 & 75.539 & 85.444 & 88.125 \\ \hline
\end{tabular}
\end{table}

\noindent\textit{\textbf{FedCAT}}: Trains the model \( M_{123} \), where clients train in parallel for tasks \( T_1 \), \( T_2 \), and \( T_3 \), followed by FL. Table \ref{tab:FedCAT_metrics} shows the performance of the final model \( M_{123} \) for tasks \( T_1 \), \( T_2 \), and \( T_3 \). However, it is evident that Fed performs optimally at different rounds for each task: round 3 for \( T_1 \), round 1 for \( T_2 \), and round 19 for \( T_3 \). We observe that additional Fed rounds degrades performance in terms of METEOR and ROUGE-L for \( T_1 \) and \( T_2 \). The final model \( M_{123} \) performs well for \( T_3 \), however performs poorly on \( T_1 \) and \( T_2 \) compared to the individual models trained in RQ1. Hence, for this method, a single BEST round could not be determined for all tasks, highlighting the instability of this multi-task training approach for FedLLM, and hence we haven't shown anecdotal examples.
\begin{table}[ht]
\centering
\sloppy
\caption{Performance Metrics for Fed Clients-And-Tasks FedLLM (in \%)}
\label{tab:FedCAT_metrics}
\begin{tabular}{|l|S[table-format=2.3]|S[table-format=2.3]|S[table-format=2.3]|S[table-format=2.3]|S[table-format=2.3]|S[table-format=2.3]|S[table-format=2.3]|S[table-format=2.3]|S[table-format=2.3]|}
\hline
\rowcolor{gray!30}
\textbf{Model} & \multicolumn{9}{c|}{\textbf{Task Metrics}} \\ \hline
\rowcolor{gray!30}
\textbf{} & \multicolumn{3}{c|}{\footnotesize \textbf{T1 (RNP)}} & \multicolumn{3}{c|}{\footnotesize \textbf{T2 (RCG)}} & \multicolumn{3}{c|}{\footnotesize \textbf{T3 (CR)}} \\ \hline
\scriptsize \textbf{BEST} & \multicolumn{3}{c|}{3} & \multicolumn{3}{c|}{1} & \multicolumn{3}{c|}{19} \\ \hline
\rowcolor{gray!30}
\textbf{} & \scriptsize \textbf{PRECISION} & \scriptsize \textbf{RECALL} & \scriptsize \textbf{F1} & \scriptsize \textbf{C-BLEU} & \scriptsize \textbf{METEOR} & \scriptsize \textbf{ROUGE-L} & \scriptsize \textbf{C-BLEU} & \scriptsize \textbf{METEOR} & \scriptsize \textbf{ROUGE-L} \\ \hline
\scriptsize \textbf{FedBEST} & 47.303 & 42.100 & 44.550 & 0.140 & 7.673 & 7.922 & 75.656 & 85.054 & 87.689 \\ \hline
\end{tabular}
\end{table}
% \begin{table}[ht]
% \centering
% \sloppy
% \caption{Performance Metrics for Fed Clients-And-Tasks FedLLM (in \%)}
% \label{tab:FedCAT_metrics}
% \begin{tabular}{|l|S[table-format=2.3]|S[table-format=2.3]|S[table-format=2.3]|S[table-format=2.3]|S[table-format=2.3]|S[table-format=2.3]|S[table-format=2.3]|S[table-format=2.3]|S[table-format=2.3]|}
% \hline
% \rowcolor{gray!30}
% \textbf{Task >} & \multicolumn{3}{c|}{\footnotesize \textbf{T1}} & \multicolumn{3}{c|}{\footnotesize \textbf{T2}} & \multicolumn{3}{c|}{\footnotesize \textbf{T3}} \\ \hline
% \scriptsize \textbf{BEST >} & \multicolumn{3}{c|}{3} & \multicolumn{3}{c|}{1} & \multicolumn{3}{c|}{19} \\ \hline
% \rowcolor{gray!30}
% \textbf{Model} & \scriptsize \textbf{PRECISION} & \scriptsize \textbf{RECALL} & \scriptsize \textbf{F1} & \scriptsize \textbf{C-BLEU} & \scriptsize \textbf{METEOR} & \scriptsize \textbf{ROUGE-L} & \scriptsize \textbf{C-BLEU} & \scriptsize \textbf{METEOR} & \scriptsize \textbf{ROUGE-L} \\ \hline
% \scriptsize \textbf{FedBEST} & 0.473 & 0.421 & 0.446 & 0.001 & 0.077 & 0.079 & 0.757 & 0.851 & 0.877 \\ \hline
% \end{tabular}
% \end{table}

\noindent\textit{\textbf{FedCFT}}: Trains the model \( M_{123} \), where clients train cumulatively for tasks \( T_1 \), \( T_2 \), and \( T_3 \), followed by FL. Table \ref{tab:FedCFT_metrics} shows the performance of the final model \( M_{123} \) for all tasks. However, it is evident that Fed performs optimally at different rounds for each task: round 2 for \( T_3 \) and round 1 for the rest. For \( T_3 \), although round 2 and 3 are best than 1, because round 1 works well for other 2 tasks, we select R1 as BEST for practicality. The final model \( M_{123} \) performs good on \( T_2 \) and \( T_3 \) compared to the individual models trained in RQ1. Anecdotal output for \( T_2 \) shown in Table \ref{tab:T2_outputs} is on the right path in addressing interface checking, but similar to FedBEST model, it misses the base class type check. Anecdotal output for \( T_3 \) is same as the ground truth , as shown in Table \ref{tab:T3_outputs}.
% <TBD> and \( T_3 \) is same as FedBEST outputs of Individual task model  and Table \ref{tab:T3_outputs} respectively. <TBD>
We observe that additional Fed rounds degrades performance, especially looses the complete ability to handle classification task T1 after round 1. This inspires the next technique where we apply FedCFT separately for classification and regression tasks. 
% The final model \( M_{123} \) performs poorly on \( T_1 \) and \( T_2 \) compared to the individual models trained in RQ1. Hence, for this method, a single BEST round could not be determined for all tasks, highlighting the instability of this multi-task training approach for FedLLM.
\begin{table}[ht]
\centering
\sloppy
\caption{Performance Metrics for Fed Cumulative Fine-Tuning FedLLM (in \%)}
\label{tab:FedCFT_metrics}
\begin{tabular}{|l|S[table-format=2.3]|S[table-format=2.3]|S[table-format=2.3]|S[table-format=2.3]|S[table-format=2.3]|S[table-format=2.3]|S[table-format=2.3]|S[table-format=2.3]|S[table-format=2.3]|}
\hline
\rowcolor{gray!30}
\textbf{Model} & \multicolumn{9}{c|}{\textbf{Task Metrics}} \\ \hline
\rowcolor{gray!30}
\textbf{} & \multicolumn{3}{c|}{\footnotesize \textbf{T1 (RNP)}} & \multicolumn{3}{c|}{\footnotesize \textbf{T2 (RCG)}} & \multicolumn{3}{c|}{\footnotesize \textbf{T3 (CR)}} \\ \hline
\scriptsize \textbf{BEST} & \multicolumn{3}{c|}{1} & \multicolumn{3}{c|}{1} & \multicolumn{3}{c|}{2} \\ \hline
\rowcolor{gray!30}
\textbf{} & \scriptsize \textbf{PRECISION} & \scriptsize \textbf{RECALL} & \scriptsize \textbf{F1} & \scriptsize \textbf{C-BLEU} & \scriptsize \textbf{METEOR} & \scriptsize \textbf{ROUGE-L} & \scriptsize \textbf{C-BLEU} & \scriptsize \textbf{METEOR} & \scriptsize \textbf{ROUGE-L} \\ \hline
\rowcolor{gray!13}
\scriptsize \textbf{FedBEST} & 79.487 & 6.200 & 11.503 & 0.766 & 9.478 & 10.952 & 75.716 & 85.126 & 87.804 \\ \hline
\scriptsize \textbf{Fed@1} & \cellcolor{gray!20}79.487 & \cellcolor{gray!20}6.200 & \cellcolor{gray!20}11.503 & \cellcolor{gray!20}0.766 & \cellcolor{gray!20}9.478 & \cellcolor{gray!20}10.952 & 75.141 & 84.760 & 87.679 \\ \hline
\scriptsize \textbf{Fed@2} & 0.000 & 0.000 & 0.000 & 0.721 & 8.015 & 10.054 & \cellcolor{gray!20}75.716 & \cellcolor{gray!20}85.126 & \cellcolor{gray!20}87.804 \\ \hline
\end{tabular}
\end{table}
% \begin{table}[ht]
% \centering
% \sloppy
% \caption{Performance Metrics for Fed Cumulative Fine-Tuning FedLLM (in \%)}
% \label{tab:FedCFT_metrics}
% \begin{tabular}{|l|S[table-format=2.3]|S[table-format=2.3]|S[table-format=2.3]|S[table-format=2.3]|S[table-format=2.3]|S[table-format=2.3]|S[table-format=2.3]|S[table-format=2.3]|S[table-format=2.3]|}
% \hline
% \rowcolor{gray!30}
% \textbf{Task >} & \multicolumn{3}{c|}{\footnotesize \textbf{T1}} & \multicolumn{3}{c|}{\footnotesize \textbf{T2}} & \multicolumn{3}{c|}{\footnotesize \textbf{T3}} \\ \hline
% \scriptsize \textbf{BEST >} & \multicolumn{3}{c|}{1} & \multicolumn{3}{c|}{1} & \multicolumn{3}{c|}{2} \\ \hline
% \rowcolor{gray!30}
% \textbf{Model} & \scriptsize \textbf{PRECISION} & \scriptsize \textbf{RECALL} & \scriptsize \textbf{F1} & \scriptsize \textbf{C-BLEU} & \scriptsize \textbf{METEOR} & \scriptsize \textbf{ROUGE-L} & \scriptsize \textbf{C-BLEU} & \scriptsize \textbf{METEOR} & \scriptsize \textbf{ROUGE-L} \\ \hline
% \rowcolor{gray!13}
% \scriptsize \textbf{FedBEST} & 0.795 & 0.062 & 0.115 & 0.008 & 0.095 & 0.110 & 0.757 & 0.851 & 0.878 \\ \hline
% \scriptsize \textbf{Fed@1} & \cellcolor{gray!20}0.795 & \cellcolor{gray!20}0.062 & \cellcolor{gray!20}0.115 & \cellcolor{gray!20}0.008 & \cellcolor{gray!20}0.095 & \cellcolor{gray!20}0.110 & 0.751 & 0.848 & 0.877 \\ \hline
% \scriptsize \textbf{Fed@2} & 0.000 & 0.000 & 0.000 & 0.007 & 0.080 & 0.101 & \cellcolor{gray!20}0.757 & \cellcolor{gray!20}0.851 & \cellcolor{gray!20}0.878 \\ \hline
% \end{tabular}
% \end{table}

\noindent\textit{\textbf{FedCFT-reg}}: Trains the regression model \( M_{23} \), where clients train cumulatively for tasks \( T_2 \), and \( T_3 \) using FedCFT technique. Trains the classification model \( M_{1} \), where clients train cumulatively for tasks \( T_1 \) using FedCFT technique. Because of a single classification task, it is same as the individual-task model trained for T1 in RQ1. Table \ref{tab:FedCFT_reg_metrics} shows the performance of the final models \( M_{1} \) on \( T_1 \), and \( M_{23} \) for the other tasks. It is evident that Fed performs optimally at different rounds for each task: round 2 for \( T_3 \) and round 1 for the rest. For \( T_3 \), although round 4 offers better ROUGE score than round 2, we choose round 2 as optimal because the other two metrics are best. For \( T_2 \), although round 1 is slightly better than round 2, we choose round 2 as BEST for practicality, because \( T_3 \) is comparatively bad in round 1. The final FedCFT-reg models perform comparable or better compared to the individual models trained in RQ1. 
Anecdotal output for \( T_2 \) shown in Table \ref{tab:T2_outputs} is comparatively closer to ground truth by addressing interface checking and considering the base class. Anecdotal output for \( T_3 \) is same as the ground truth, as shown in Table \ref{tab:T3_outputs}.
\begin{table}[ht]
\centering
\sloppy
\caption{Performance Metrics for Fed Cumulative Fine-Tuning - Separate for Regression FedLLM (in \%)}
\label{tab:FedCFT_reg_metrics}
\begin{tabular}{|l|S[table-format=2.3]|S[table-format=2.3]|S[table-format=2.3]|S[table-format=2.3]|S[table-format=2.3]|S[table-format=2.3]|S[table-format=2.3]|S[table-format=2.3]|S[table-format=2.3]|}
\hline
\rowcolor{gray!30}
\textbf{Model} & \multicolumn{9}{c|}{\textbf{Task Metrics}} \\ \hline
\rowcolor{gray!30}
\textbf{} & \multicolumn{3}{c|}{\footnotesize \textbf{T1 (RNP)}} & \multicolumn{3}{c|}{\footnotesize \textbf{T2 (RCG)}} & \multicolumn{3}{c|}{\footnotesize \textbf{T3 (CR)}} \\ \hline
\scriptsize \textbf{BEST} & \multicolumn{3}{c|}{\cellcolor{blue!10}{1}} & \multicolumn{3}{c|}{1} & \multicolumn{3}{c|}{2} \\ \hline
\rowcolor{gray!30}
\textbf{} & \scriptsize \textbf{PRECISION} & \scriptsize \textbf{RECALL} & \scriptsize \textbf{F1} & \scriptsize \textbf{C-BLEU} & \scriptsize \textbf{METEOR} & \scriptsize \textbf{ROUGE-L} & \scriptsize \textbf{C-BLEU} & \scriptsize \textbf{METEOR} & \scriptsize \textbf{ROUGE-L} \\ \hline
\scriptsize \textbf{FedBEST} & \cellcolor{blue!15}49.141 & \cellcolor{blue!15}62.900 & \cellcolor{blue!15}55.175 & \cellcolor{gray!13}0.667 & \cellcolor{gray!13}9.323 & \cellcolor{gray!13}10.921 & \cellcolor{gray!13}76.104 & \cellcolor{gray!13}85.560 & \cellcolor{gray!13}88.142 \\ \hline
\scriptsize \textbf{Fed@1} & \cellcolor{blue!15}49.141 & \cellcolor{blue!15}62.900 & \cellcolor{blue!15}55.175 & \cellcolor{gray!13}0.667 & \cellcolor{gray!13}9.323 & \cellcolor{gray!13}10.921 & 75.313 & 84.832 & 87.985 \\ \hline
\scriptsize \textbf{Fed@2} & \cellcolor{blue!10}54.121 & \cellcolor{blue!10}39.400 & \cellcolor{blue!10}45.602 & 0.861 & 8.419 & 10.552 & \cellcolor{gray!13}76.104 & \cellcolor{gray!13}85.560 & \cellcolor{gray!13}88.142 \\ \hline
\end{tabular}
\end{table}

Table \ref{tab:CodeReviewAutomation_metrics} presents a comparison of all the models discussed in this study, including vanilla, individual-task models, and multi-task models. Among the individual-task models, the FedBEST model performs comparably to the centrally trained model, both significantly improving the performance over the vanilla model.  Using the Wilcoxon signed-rank test \cite{woolson2007wilcoxon}, we observe that the Central and FedBEST models significantly outperform the Vanilla model across the evaluated metrics (\(p = 0.027\) that is p-value is less than 0.05 for both comparisons), indicating that the performance improvements achieved by individual-task fine-tuning are statistically significant.
% Null hypothesis: Individual models' performance is significantly better than vanilla model.

Among the multi-task models, CFT shows slightly better performance than CFT-reg for \(T_2\), while CFT-reg outperforms for \(T_3\). The results indicate that an individual model should be used for \(T_1\) in both CFT and CFT-reg. Given that CFT-reg requires fewer training samples for the regression model (as it excludes the training data from \(T_1\)), we identify CFT-reg as the best-performing multi-task technique for code review automation. We infer that, for our dataset and experimental setup, the multi-task model leveraging task relatedness performs comparably or better than the individual-task models. A two-sided Wilcoxon signed-rank test \cite{woolson2007wilcoxon} confirms that the multi-task CFT-reg model significantly outperforms the individual-task FedBEST model across the evaluated metrics (\(p = 0.0277\)), demonstrating that leveraging task relatedness provides measurable performance gains over individual-task fine-tuning.

% We infer that, for our dataset and experimental setup, the multi-task model leveraging task relatedness performs comparably or better than the individual-task models. Using a two-sided Wilcoxon signed-rank test, we find that the multi-task model CFT-reg significantly outperforms the individual-task FedBEST model across the evaluated metrics (\(p = 0.0277\) that is p-value is less than 0.05), demonstrating that leveraging task relatedness provides measurable performance gains over individual-task fine-tuning.
% Null hypothesis: There is no significant difference between the performance of the FedBEST and CFT-reg models. Alternative Hypothesis: The performance of the CFT-reg model is different (better or worse) than that of the FedBEST model. Test statistic of 0.0 suggests that CFT-reg consistently outperforms FedBEST
\begin{table}[ht]
\centering
\small
\sloppy
\caption{Performance Metrics for Various Models in Code Review Automation (in \%).}
\label{tab:CodeReviewAutomation_metrics}
\begin{tabular}{|p{1.2cm}|S[table-format=2.3]|S[table-format=2.3]|S[table-format=2.3]|S[table-format=2.3]|S[table-format=2.3]|S[table-format=2.3]|S[table-format=2.3]|S[table-format=2.3]|S[table-format=2.3]|}
\hline
\rowcolor{gray!30}
\textbf{Model} & \multicolumn{9}{c|}{\textbf{Task Metrics}} \\ \hline
\rowcolor{gray!30}
\textbf{} & \multicolumn{3}{c|}{\footnotesize \textbf{T1 (RNP)}} & \multicolumn{3}{c|}{\footnotesize \textbf{T2 (RCG)}} & \multicolumn{3}{c|}{\footnotesize \textbf{T3 (CR)}} \\ \hline
\textbf{}  & \scriptsize \textbf{PRECISION} & \scriptsize \textbf{RECALL} & \scriptsize \textbf{F1} & \scriptsize \textbf{C-BLEU} & \scriptsize \textbf{METEOR} & \scriptsize \textbf{ROUGE-L} & \scriptsize \textbf{C-BLEU} & \scriptsize \textbf{METEOR} & \scriptsize \textbf{ROUGE-L} \\ \hline
\footnotesize \textbf{Vanilla} & 53.989 & 29.100 & 37.817 & 0.110 & 8.840 & 7.706 & 12.607 & 44.878 & 34.451 \\ \hline
\footnotesize \textbf{Central} & 48.851 & 63.800 & 55.334 & 0.568 & 9.624 & 11.389 & 74.749 & 84.336 & 88.394 \\ \hline
\rowcolor{yellow!10}
% Individual 
\footnotesize \textbf{FedBEST} & 49.141 & 62.900 & 55.175 & 0.612 & 9.245 & 10.850 & 74.583 & 84.542 & 87.472 \\ \hline
\footnotesize \textbf{TOC} & 0.000 & 0.000 & 0.000 & 0.315 & 2.927 & 3.179 & 74.883 & 84.840 & 87.427 \\ \hline
\footnotesize \textbf{COT} & 39.450 & 4.300 & 7.755 & 0.216 & 3.424 & 3.325 & 75.539 & 85.444 & 88.125 \\ \hline
\footnotesize \textbf{CAT} & 47.303 & 42.100 & 44.550 & 0.140 & 7.673 & 7.922 & 75.656 & 85.054 & 87.689 \\ \hline
\footnotesize \textbf{CFT} & 79.487 & 6.200 & 11.503 & 0.766 & 9.478 & 10.952 & 75.716 & 85.126 & 87.804 \\ \hline
\rowcolor{green!20}
\footnotesize \textbf{CFT-reg} & 49.141 & 62.900 & 55.175 & 0.667 & 9.323 & 10.921 & 76.104 & 85.560 & 88.142 \\ \hline
\end{tabular}
\end{table}

\vspace{5pt}
\noindent\colorbox{yellow!50!black}{\parbox{\dimexpr\linewidth-2\fboxsep}{\color{yellow!10!white}\textbf{RQ2 Summary}}}
\noindent\fcolorbox{yellow!50!black}{yellow!10!white}{%
    \parbox{\dimexpr\linewidth-2\fboxsep-2\fboxrule\relax}{%
        \strut This study aims to develop a multi-task FedLLM for code review automation tasks by taking advantage of the relatedness between these tasks. The findings indicate that sequential training of a FedLLM on these tasks results in catastrophic forgetting, making it an unstable approach. In contrast, the cumulative fine-tuning technique for multi-task FedLLM shows improved performance over individual-task models for code review automation.
        % This study aims to develop a multi-task FedLLM for tasks in Code Review Automation, by taking advantage of the relatedness between the tasks. The findings indicate that sequential training of a FedLLM for these tasks leads to catastrophic forgetting, making it an unstable approach. However, the cumulative fine-tuning technique towards multi-task FedLLM demonstrates better performance than individual-task models for the code review automation tasks.
        % This study aims to develop a multi-task FedLLM for diverse tasks in Code Review Automation, including classification, natural and programming language generation.% However, the results demonstrate that pre-training on prior tasks might enhance the performance of subsequent tasks. Future research should explore methods to successfully build multi-task FedLLMs.
    }%
}
\vspace{5pt}
\section{Discussion and Implication}
\label{sec: Discuss}

% \noindent\textit{\textbf{Discussion:}} 
\noindent In RQ1, the tasks are treated independently, without considering their interrelated nature. We believe the connectedness of sub-tasks could be advantageous in multi-task training, and private participants could try to build a single collaborative model together which can handle all the connected tasks robustly for all the clients. So, RQ2 treats the three tasks as related, hypothesizing that the multi-task model in RQ2 will outperform individual models due to the additional context. 
However, our experimental results indicate that achieving comparable or superior performance with a multi-task model, compared to individual models, requires careful design and implementation of the multi-task approach.
The potential reasons for this are as follows:  (i) the multi-task model might be suffering from information loss that is catastrophic forgetting (CF) due to multiple retraining (T1 then T2 then T3) of the LLM. (ii) the classification task in task 1 differs significantly from the regression tasks in tasks 2 and 3, this difference serves as a basis for FedCFT-reg technique.
(iii) It could not even be our proposed approaches, the FL approach itself includes re-training the model multiple times which could be the reason for CF.  
(iv) We are training the model with (26k x 3) unique rows, however the individual models and base paper fine-tunes a model on maximum of 26k data rows. Might be training on (26k / 3) = 8666 rows per task with a total of 26k data points will give better results. 
Addressing these issues in future research could improve the performance of multi-task FedLLMs.

SE studies \cite{weyssow2023usage, gao2023keeping} have observed catastrophic forgetting (CF) in smaller language models such as CodeBERT, CodeT5, GPT-2, and RoBERTa when training for an SE task. Recent ML studies \cite{zhang2024dissecting, zhai2024investigating} have also observed CF in LLMs when trained on multiple tasks. Future research could deep-dive and experiment with continual learning techniques such as 
% cumulative fine-tuning, 
replay-based \cite{weyssow2023usage}, regularization-based \cite{weyssow2023usage}, and Dirichlet continual learning \cite{zengdirichlet} to mitigate or reduce CF when training multi-task FedLLM. 
\\
\\

\noindent\textit{\textbf{Potential Implication:}} 
\begin{enumerate}[label=\roman*)]
    \item Closed-source participants can engage in collaborative training for multi-task modeling, such as automating code reviews. SE practices advocate for decomposing tasks into manageable multi-task frameworks, making this research relevant to a variety of SE tasks.
    \item This study demonstrates that collaboration among closed-source clients can enhance model performance compared to individual client models. This approach opens promising avenues for privacy-aware academic partnerships with industry, facilitating access to proprietary data.
    \item Future research should explore continuous training scenarios in code review automation, focusing on project and task timelines. 
\end{enumerate}

\section{Threats to Validity}
\label{sec: Threats}
\textbf{External validity: }
This study focused on fine-tuning a FedLLM for code review automation using a specific GitHub dataset. The uniqueness of this dataset may limit the generalizability of our findings. Different datasets from other platforms or proprietary projects might yield different results, however in this study, we choose a multi-language dataset to ensure a broad evaluation. Additionally, this is the first SE study which assess the effectiveness of the multi-task FedLLM for code review automation. Further research using varied models and datasets is needed to validate and generalize these findings comprehensively.

\noindent\textbf{Internal validity: } 
The training in this study is limited by factors such as the training dataset (provides a single ground truth for reference, although code or comments can be written in multiple valid ways), number of epochs, model choice, and fine-tuning techniques. Different datasets and metrics might influence the optimal number of federated rounds. Further empirical analysis and research are required to build an efficient multi-task model.

\noindent\textbf{Reliability: } 
Reliability checks the method consistency when experiments are repeated with the same data. There are no previous studies on multi-task FedLLM training for code review automation to validate our results. The results are based on our implementation of FedLLM using LoRA. Future research to replicate on different federated architectures, as variations in design might impact the experiment's efficiency and outcomes.

\noindent\textbf{Construct validity: }
The effectiveness of the model relies on the appropriate setting of hyper-parameters. To ensure the selected LLM is optimally used for code review automation, we experimented to chose rank and target modules, and aligned the other hyper-parameters with literature. 
% The diverse nature of tasks at hand needs to be considered in further multi-task FedLLM research. 

\section{Related Work}
\label{sec: Related}
\noindent\textbf{Code Review Automation:} 
The code review process is critical for validating code before deployment in production, helping to identify bugs and improve both functional and non-functional qualities of the code \cite{tufano2024code, bacchelli2013expectations}. Additionally, it offers benefits such as knowledge transfer, exposure to alternative thought processes, and increased team awareness \cite{sadowski2018modern, bacchelli2013expectations, sripada2015support}. However, due to its time-consuming nature, researchers have been exploring ways to automate parts of the process, such as assigning reviewers \cite{al2020workload, balachandran2013reducing}. More recently, multiple aspects of the code review process have been automated, including mimicking human reviewers to generate code review comments \cite{tufano2024code, tufano2021towards}. Existing research \cite{lu2023llama, li2022automating} approaches the task through three primary sub-tasks: (i) code review necessity prediction (T1), (ii) code review comment generation (T2), and (iii) code refinement based on review comments (T3). Our paper adopts a similar approach, tackling the modern code review automation process with these three sub-tasks.

\noindent\textbf{Models for Code Review Automation:} 
A study \cite{tufano2021towards} has approached code review automation using transformer-based encoder-decoder deep learning (DL) models. For T2, one encoder is used to encode code changes, and one decoder generates the comments. For T3, two encoders are used to encode both code changes and comments, with one decoder generating the refined code. A study \cite{tufano2022using} employed the pre-trained T5 model for T2 and T3, demonstrating higher performance compared to traditional DL encoder-decoder models. The CodeReviewer \cite{li2022automating} paper, which introduced the T1, T2, and T3 tasks followed in this study, further pre-trained a CodeT5-based model for tasks including denoising code diffs, denoising code comments, diff tag prediction, and code comment generation, enhancing its capabilities in T1, T2, and T3. Studies \cite{li2022automating} have also utilized the CodeT5 model, showing that T5 and CodeT5 models perform best for T2 and T3, respectively.

LLMs are huge pre-trained models capable of handling generic tasks and can be further fine-tuned to specialize in specific tasks or domains, hence have been used for SE research \cite{lu2023llama, pornprasit2024fine}. LLaMA-Reviewer \cite{lu2023llama} fine-tunes an LLM, Llama2, using Zero-init Attention Prefix-tuning and Low-Rank Adaptation (LoRA) to train for T1, T2, and T3, demonstrating that the combination of LoRA \cite{hu2021lora} and LLaMA \cite{touvron2023llama} is effective for code review automation. This paper serves as the basis for the LoRA fine-tuning approach for our study. However, we extend this approach by training the model using Federated Learning (FL), which allows the data to remain private.

\noindent\textbf{Federated Learning in SE:} 
Recently, FL has been increasingly utilized in SE studies. A study \cite{shanbhag2022exploring} used FL to train a DL model for labeling bug-fix commits, where a pre-trained BERT model with 110M parameters was trained on less than 2k data points using FL with three clients. Research on FL architectural patterns \cite{lo2022architectural, zhang2020federated} has studied the trade-off between communication latency, model evolution time, and performance across different FL architectures, including centralized, hierarchical, regional, and decentralized. The results indicated that although centrally trained models have higher latency and evolution time, their performance is superior for gaining generic domain or task knowledge \cite{zhang2020federated}. Hence, we adopt a central architecture in this paper.

A literature review \cite{smestad2023systematic} on client selection for FL suggests that, in real-world scenarios, participants should be selected by considering heterogeneity, communication cost, resources, and fairness. In our study, all clients have heterogeneous data but are fine-tuning for the same tasks. Our study implements a simulated environment, but these factors should be considered in practical applications.
A recent study \cite{yang2024federated} employed SVM and CodeBERT-based models for code clone detection and defect prediction tasks. While all the above discussed studies utilized comparatively smaller language models, a more recent study \cite{kumar2024code} trained a large language model, LLaMA-based using LoRA for the code summarization task. This paper serves as the basis for the federated architecture we use in our study. To our knowledge, ours is the first work that utilizes FedLLM and attempts to train a multi-task FedLLM for code review automation tasks.

\section{Conclusion and Future Directions}
\label{sec: ConclFuture}
This study investigates the development of a FedLLM for automating code review tasks, which includes review necessity prediction (RNP), review comment generation (RCG), and code refinement (CR). By employing federated fine-tuning on the 8B LLaMA-3 model, we found that while individual FedLLMs for RNP and RCG primarily benefit low-data client, the CR FedLLM enhances performance for both high-data and low-data clients. Our results indicate that the FedLLM outperforms the vanilla pre-trained LLM, achieving an 18\% improvement in F1 score for RNP, a 3\% increase in ROUGE-L for RCG, and an impressive 53\% enhancement in ROUGE-L for CR. 

Furthermore, to leverage the interrelated nature of these code review tasks, we attempted to construct a multi-task FedLLM capable of addressing all three tasks simultaneously. Our findings revealed that sequential multi-task training resulted in catastrophic forgetting, making it less effective than training individual models for each task. Whereas, the cumulative training technique demonstrated better performance compared to the individual-task models, enabling the deployment of a single FedLLM that can effectively manage related tasks. 

Future research should focus on addressing the limitations identified in this study to enhance the performance of multi-task federated LLMs for code review automation. Potential areas for exploration include implementing continual learning techniques such as reply-based or regularization techniques to mitigate catastrophic forgetting, as well as investigating alternative training approaches that target different modules for specific tasks, which may improve the stability and performance of multi-task FedLLMs in SE task. Additionally, future studies could explore the integration of privacy-preserving techniques, such as homomorphic encryption, to strengthen the security of collaborative FedLLM modeling. 
% The classic transfer learning attempted highlights the challenges in building efficient multi-task FedLLMs, and also shows the need for further research in the continual fine-tuning techniques in SE research. The cumulative techniques proved to be best among the multi-task FedLLM techniques tried and is offering better performance than the individual models, hence showing the need for further research to utilize the relatedness between the sub-tasks in code review automation.  

% Future research could address the limitations identified in this study to improve the performance of multi-task FedLLMs for code review automation. Potential areas for exploration include experimenting with reply-based, or regularization-based techniques to mitigate catastrophic forgetting. Exploring alternative training approaches further, such as targeting different modules for different tasks may enhance the stability and performance of multi-task FedLLMs for SE tasks. Future research could investigate the integration of additional privacy-preserving techniques, such as homomorphic encryption, to enhance the security of collaborative FedLLM modeling. 

\section{Data Availability}
\label{sec: DataAvail}
Artifacts of this study are made available. \footnote{Replication package: \url{https://osf.io/ztnhk/?view_only=b5d82d817e87461abfc218cefde60b0c }}
% <TBD> 
\section{ACKNOWLEDGMENTS}
\label{sec: Ack}
% The authors are thankful to the anonymous reviewers for their constructive comments. 
We acknowledge support from the Department of Computer Science and Engineering for providing access to a High Performance Computing facility based on Nvidia DGX equipment.
  
% \balance
\bibliographystyle{ACM-Reference-Format}
\bibliography{references}
\end{document}